\documentclass[11pt,a4paper]{article}

\usepackage{a4}          %

\usepackage[UKenglish]{babel}
\usepackage{amsmath}
\usepackage{amssymb}
\usepackage[usenames]{color} 
\usepackage{multirow}
\usepackage{graphicx}
\usepackage{epstopdf}

\usepackage{ifpdf}
\ifpdf
  \usepackage[pdftex,
    pdftitle={U(1) symmetries  in F-theory GUTs with multiple sections},
    pdfauthor={Christoph Mayrhofer, Eran Palti, Timo Weigand},
    pdfsubject={F-theory compactifications},
    bookmarksopen, bookmarksnumbered, bookmarksopenlevel=2]{hyperref}
\fi
\def\hybrid{\topmargin -20pt    \oddsidemargin 0pt
        \headheight 0pt \headsep 0pt
        \textwidth 6.25in       %
        \textheight 9 in       %
        \marginparwidth .875in
        \parskip 5pt plus 1pt 
          \jot = 1.5ex
   }
\hybrid
\numberwithin{equation}{section}
\numberwithin{table}{section}

\newcommand{\bal}{\begin{aligned}}   \newcommand{\eal}{\end{aligned}}

\newcommand{\bbP}{\mathbb{P}}

\newcommand{\pl}[1]{\bbP^1_{\,#1}}

\newcommand{\nn}{\nonumber}

\newcommand{\C}{\text{C}}
\newcommand{\PP}{\text{P}}
\newcommand{\Q}{\text{Q}}
\newcommand{\B}{\text{B}}
\newcommand{\E}{\text{E}}
\newcommand{\aK}{\bar{K}}

\newcommand{\tw}{\text{w}}
\newcommand{\be}{\begin{equation}}

\newcommand{\executeiffilenewer}[3]{%
 \ifnum\pdfstrcmp{\pdffilemoddate{#1}}%
 {\pdffilemoddate{#2}}>0%
 {\immediate\write18{#3}}\fi%
}

\begin{document}

\baselineskip=14pt
\parskip 5pt plus 1pt

\vspace*{-1.5cm}
\begin{flushright}    %
  {\small
 CPHT-RR087.1112
  }
\end{flushright}

\vspace{2cm}
\begin{center}        %
  {\LARGE $U(1)$ symmetries  in F-theory GUTs with multiple sections}
\end{center}

\vspace{0.75cm}
\begin{center}        %
 Christoph Mayrhofer$^{1}$, Eran Palti$^{2}$, Timo Weigand$^{1}$
\end{center}

\vspace{0.15cm}
\begin{center}        %
  \emph{$^{1}$ Institut f\"ur Theoretische Physik, Ruprecht-Karls-Universit\"at, \\
             Heidelberg, Germany}
             \\[0.15cm]
  \emph{$^{2}$ Centre de Physique Theorique, Ecole Polytechnique,\\ 
               CNRS, %
               Palaiseau, France
  }
\end{center}

\vspace{2cm}

\begin{abstract}

We present a systematic construction of F-theory compactifications with Abelian gauge symmetries in addition to a non-Abelian gauge group $G$. The formalism is generally applicable to models in global Tate form but we focus on the phenomenologically interesting case of $G=SU(5)$. The Abelian gauge factors arise due to extra global sections resulting from a specific factorisation of the Tate polynomial which describes the elliptic fibration. These constructions, which accommodate up to four different $U(1)$ factors, are worked out in detail for the two possible embeddings of a single $U(1)$ factor into $E_8$, usually denoted $SU(5) \times U(1)_X$ and $SU(5) \times U(1)_{PQ}$. 
The resolved models can be understood either patchwise via a small resolution or in terms of a $\mathbb P_{1,1,2}[4]$ description of the elliptic fibration. We derive the $U(1)$ charges of the fields from the geometry, construct the $U(1)$ gauge fluxes and exemplify the structure of the Yukawa interaction points. 
A particularly interesting result is that the global $SU(5) \times U(1)_{PQ}$ model exhibits extra $SU(5)$-singlet states which are incompatible with a single global decomposition of the ${\bf 248}$ of $E_8$. The states in turn lead to new Yukawa type couplings which have not been considered in local model building.

\end{abstract}

\clearpage

\newpage

\tableofcontents

\section{Introduction}

The exceptional group $E_8$ plays a central role in string theory and in particular in the Heterotic string and F-theory. Within the context of F-theory Grand Unified Theories (GUTs) {\cite{Beasley:2008dc} an underlying $E_8$ implies that the GUT group, minimally $SU(5)$, is naturally extended by additional symmetries coming from its embedding in $E_8$. Particularly interesting for model building are additional $U(1)$ symmetries. These play two important roles: they can support gauge flux thereby inducing chirality in the massless spectrum, and they can lead to gauge and global symmetries beyond the Standard Model that can be used to control the structure of the low energy theory, e.g.\ to forbid proton decay or generating flavour hierarchies. In F-theory \cite{Vafa:1996xn,Morrison:1996na} gauge symmetries are geometric in origin and so understanding the geometry associated to $U(1)$ symmetries is crucial to this aspect of model building. 
Indeed a lot of recent progress has been made towards understanding the explicit construction and fully global aspects of $U(1)$s in four-dimensional F-theory compactifications \cite{Grimm:2010ez,Braun:2011zm,Krause:2011xj,Grimm:2011fx,Krause:2012yh,Marsano:2012yc,Cvetic:2012xn} and in compactifications to six dimensions \cite{Morrison:2012ei}.  In the local approach to F-theory $U(1)$ symmetries are quite well understood within the spectral cover approach \cite{Donagi:2009ra} and have been used extensively in local model building \cite{Marsano:2009gv,Marsano:2009wr,Dolan:2011iu,Palti:2012dd,Dudas:2010zb,Dudas:2009hu}. However, many important aspects of $U(1)$ symmetries are inherently global in nature: they can be broken away from the GUT brane \cite{Hayashi:2010zp,Grimm:2010ez} and the associated gauge flux is not localised on the GUT brane. Therefore a global understanding of $U(1)$ symmetries is one of the central requirements for realistic F-theory model building. 

In the weakly coupled type IIB limit it is possible to study intersecting D7-brane configurations generally without specifying the explicit geometry of the Calabi-Yau three-fold. The analogous general procedure for F-theory models is the study of the elliptic fibration without specifying the base explicitly. The form that the elliptic fibration must take in order to induce a non-Abelian singularity, or equivalently gauge group, over a divisor in the base is very well understood and given a fibration the non-Abelian structure can be discerned in an algorithmic way \cite{Bershadsky:1996nh,Katz:2011qp}. Further, the $U(1)$ components that make up the Cartan of a non-Abelian singularity can be studied explicitly by considering the M-theory dual. On the M-theory side it is possible to resolve the non-Abelian singularity, which in the gauge theory corresponds to moving along the Coulomb branch. After resolving the singularity each Cartan element corresponds to a resolution divisor whose dual two-form gives rise to 
a $U(1)$ gauge field from dimensional reduction of the M-theory three-form $C_3$. In the context of four-dimensional F-theory $SU(5)$ GUT models this procedure has been carried out, using various techniques, in 
\cite{Blumenhagen:2009yv,Esole:2011sm,Marsano:2011hv,Krause:2011xj,Grimm:2011fx,Krause:2012yh} (see  \cite{Chen:2010ts} for other gauge groups).

Abelian symmetries that are not in the Cartan of a non-Abelian singularity are less well understood. The first complication is that in string theory such isolated $U(1)$ symmetries can often gain a St\"uckelberg mass removing them from the massless spectrum. In the weakly coupled IIB limit this is possible even in the absence of any flux, and such a purely geometrically massive $U(1)$ would be very difficult to identify in the F-theory uplift. Some progress towards understanding such $U(1)$s was made in \cite{Grimm:2010ez,Grimm:2011tb} where they were proposed to uplift to non-closed two-forms on the M-theory side. Further in \cite{Krause:2012yh} the flux associated to one such massive $U(1)$, the diagonal one in the IIB limit, was identified. However a general procedure for identifying and constructing such $U(1)$ symmetries is missing and we will have nothing new to say regarding them in this paper. 

A more tractable class of $U(1)$ symmetries are those which remain massless in the absence of any flux. 
A general approach to such $U(1)$ symmetries should involve the construction of multiple sections of the elliptic fibration. It was shown already in \cite{Morrison:1996na} that isolated $U(1)$s correspond to additional sections beyond the universal one which specifies %
the embedding of the base.
Concrete investigations of $U(1)$ symmetries in six-dimensional F-theory compactifications have appeared early on in \cite{Aldazabal:1996du}.
In the context of $SU(5)$ GUTs one approach to realising massless $U(1)$s was proposed in \cite{Grimm:2010ez} in terms of what was called an $U(1)$-restricted Tate model.
The idea was to choose the coefficients of the elliptic fibration, which already have an $SU(5)$ singularity over the GUT divisor, so as to induce an additional $SU(2)$ singularity in the fibre over a curve in the base. The resolution of this singularity introduces a new divisor  which is associated to the additional $U(1)$ symmetry \cite{Grimm:2010ez,Krause:2011xj,Grimm:2011fx,Krause:2012yh}. In \cite{Braun:2011zm} it was shown that the $SU(2)$ singularity can be written in the form of a conifold which is then resolved. The same procedure was used \cite{Marsano:2012yc} to construct models with an additional $U(1)$ symmetry. The $U(1)$-restricted Tate model is quite well understood by now. However this model realises only one particular embedding of a single additional $U(1)$ symmetry in $E_8$ and additionally strongly restricts the possible matter spectrum, by turning off one of the {\bf 10}-matter curves, from the most general configuration. This is in contrast to the rich structure of $U(1)$ symmetries 
and matter spectra that are possible in breaking $E_8 \rightarrow SU(5)$ which have been used in local model building \cite{Marsano:2009gv,Marsano:2009wr,Dolan:2011iu,Palti:2012dd,Dudas:2010zb,Dudas:2009hu}. Indeed in its local limit, i.e.\ in the projection to the $SU(5)$ GUT divisor, the $U(1)$ restricted Tate model flows to the $SU(5) \times U(1)_X$ split spectral cover \cite{Donagi:2009ra,Marsano:2009gv}. Locally, one usually thinks of the gauge group $SU(5)$ as arising from an underlying $E_8$ symmetry and the possible $U(1)$s then arise from the various embeddings into $E_8$ \cite{Dudas:2010zb,Dolan:2011iu}. The purpose of this paper is to study how the full spectrum of possibilities can be realised in a global setting thereby opening the way to realising the phenomenology of local models in a global string vacuum. This route will offer some surprises.

The key idea of our present paper is that we construct Tate models which give in a specific way multiple sections. %
We call these factorised Tate models. We will show that such models automatically induce a binomial singularity on the manifold whose resolution gives rise to the appropriate $U(1)$s and their fluxes. 
The importance of suitable global factorisations of the spectral cover equation to construct heterotic models with (multiple) $U(1)$ symmetries has been recently explored in \cite{Choi:2012pr}. Our approach is independent of any heterotic dual.

By explicitly resolving and studying the fibre structure in detail for some examples we derive the matter spectrum and the associated $U(1)$ charges directly from the geometry.
A crucial aspect of the construction is that, unlike in the $U(1)$-restricted model, we will recover the full matter spectrum with no additional constraints on the matter curves.
We work out the details of these matter curves and their resolved fibre structure both for models with so-called $U(1)_X$ charge and with $U(1)_{PQ}$ charge. The latter has been used intensively in local model building because the Peccei-Quinn symmetry can solve the $\mu$-problem and forbid dimension-five proton decay operators \cite{Marsano:2009wr}, and our analysis provides the first global embedding of this scenario.

The spectrum of $SU(5)$ charged matter indeed assembles into representations that can be obtained by the decomposition of the adjoint of one $E_8$ into $SU(5)$. In this sense factorised Tate models are the appropriate way to systematically construct the fibrations that account for the spectrum of possible embeddings of $U(1)$ symmetries arising from breaking $E_8$ to $SU(5)$. Most surprisingly, however,  this structure is in general not respected by the $SU(5)$ singlet states charged under the $U(1)$. Indeed we will exemplify for the $SU(5) \times U(1)_{PQ}$ model that, contrary to expectations based on local reasoning, the entire spectrum does not assemble into a single $E_8$ representation once we take all singlets into account. Since the singlets are localised on curves away from the GUT divisor, such behaviour can only be detected in a global approach. 
 However, the singlet curves do intersect the GUT divisor in points at which couplings to the $SU(5)$ charged matter are localised, and they are thus also relevant for local model building and phenomenology. To the extent that the most general pattern of Yukawa couplings cannot, as we find, be unified in a single $E_8$-point, the idea of a general underlying $E_8$ symmetry present along the entire GUT divisor cannot be maintained.  

Another crucial aspect of model building in F-theory is background gauge flux and this is intimately related with extra $U(1)$ symmetries as well. Within a global setting our current understanding of gauge flux requires a fully resolved and smooth manifold and is realised on the M-theory side as the 4-form $G_4$-flux. For pure $SU(5)$ models $G_4$-flux has been studied in \cite{Marsano:2010ix, Marsano:2011hv}. This flux was identified in \cite{Krause:2012yh} as a massive $U(1)$ flux from a Type IIB perspective.\footnote{See the last two references in \cite{Chen:2010ts}  for the analogous fluxes in $E_6$ and $SO(10)$ models, respectively.} In the presence of massless $U(1)$ symmetries, the associated $G_4$-flux was studied in \cite{Braun:2011zm} for a $U(1)$ model, and in \cite{Krause:2011xj,Grimm:2011fx,Krause:2012yh} for an $SU(5)\times U(1)$ model based on the $U(1)$-restricted Tate model. Crucial to the definition of the flux is the full resolution of all the singularities as the expression of $G_4$ 
involves the resolution divisors from both the $SU(5)$ singularity and the additional $SU(2)$ singularity. Since one of the primary motivations for constructing additional $U(1)$ symmetries is that the flux associated to them can induce chirality in the visible sector, in this paper we also present a construction of the associated $G_4$-flux.

Our construction fits nicely into the approach of \cite{Morrison:2012ei}, which gives the general form of the Weierstra\ss{} equation for an elliptic fibration with two sections (and therefore one non-Cartan $U(1)$), but which is otherwise generic, i.e.\ has a priori no non-Abelian gauge symmetries built in. This singular Weierstra\ss{} model is resolved in \cite{Morrison:2012ei} and described by a smooth fibration with $\mathbb P_{1,1,2}[4]$ fibre. We show that the factorised Tate models corresponding to $SU(5) \times U(1)_X$  and $SU(5) \times U(1)_{PQ}$ can be mapped to a specialisation (due to the extra $SU(5)$ symmetry) of the model of \cite{Morrison:2012ei} which is particularly useful for studying the $U(1)$ charged singlets. In fact the appearance of a rich pattern of such singlets had been observed already in \cite{Morrison:2012ei}, albeit in a different context.

The paper is set out as follows. In section \ref{sec:factate} we introduce factorised Tate models and identify the appropriate sections and singularities corresponding to the $U(1)$ symmetries. We then explicitly resolve the two possible types of  $SU(5) \times U(1)$ Tate models presenting the resolved manifold patchwise. In sections \ref{sec:fibstruc-gen} and \ref{sec:fibu1pq} we proceed to analyse in detail the fibre structure of these $SU(5) \times U(1)_X$ and $SU(5) \times U(1)_{PQ}$ models and derive the matter spectrum. We work out the fibre structure over a selection of Yukawa points and in particular present the form over the point corresponding to a ${\bf 1\; 10\; \overline{10}}$ coupling. We also present the $G_4$-flux associated to the $U(1)$ symmetries constructing the associated two-form through the Shioda map \cite{Shioda1}. In section \ref{sec:relmorpa} we study the map between the factorised Tate models with a single $U(1)$ and the general two-section models of \cite{Morrison:2012ei}. This in 
particular confirms the presence of a novel type of singlets in the $U(1)_{PQ}$ model which does not fit into the pattern of a single underlying $E_8$. We summarise our results in section \ref{sec:disc}. In appendix \ref{sec:morfactate} we present the factorised Tate models for the other possible embeddings of $U(1)$s in $E_8$ including multiple $U(1)$s up to the maximum four. In appendix \ref{sec:relother} we explain the relation between our approach and the $U(1)$-restricted Tate model as well as the local split spectral cover.

\section{\texorpdfstring{$U(1)$}{U(1)} symmetries from the factorised Tate model}
\label{sec:factate}

\subsection{Engineering extra sections by factorisation}
\label{sec_factorisation}

F-theory compactifications to four dimensions are defined in terms of an elliptically fibred Calabi-Yau 4-fold $Y_4: T^2 \rightarrow B$ with a section that specifies the base $B$ as a submanifold of $Y_4$. This universal, so-called zero section is required so as to interpret $B$ as the physical compactification space. In F-theory massless (up to $G_4$-flux induced St\"uckelberg masses) $U(1)$ symmetries are counted by the number of additional sections \cite{Morrison:1996na}. A section specifies a point in the torus over every point in the base. We write the torus as the Weierstra\ss{} equation 
\begin{equation}
y^2 = x^3 + f x z^4 + g z^6 \;
\end{equation} 
in weighted projective space ${\mathbb P}_{[2,3,1]}$ with homogeneous coordinates $\left[x,y,z\right]$. A section is now specified by two holomorphic equations in $\left[x,y,z\right]$ whose intersection lies on the torus. For the special case of the zero section the two polynomials are $z=0$ and the Weierstra\ss{} equation. In this paper we are interested in a specific set of sections that are defined for cases where the elliptic fibration can be written in the Tate form
\begin{equation}
y^2 = x^3 + a_1 x y z + a_2 x^2 z^2 + a_3 y z^3 + a_4 x z^4 + a_6 z^6 \;. \label{tateform}
\end{equation} 
This defines the Calabi-Yau 4-fold $Y_4$ as a hypersurface in an ambient 5-fold $ X_5$. It is generally not always possible to write the elliptic fibration in this way while retaining the holomorphicity of the $a_i$, but in the case of $SU(5)$ GUT models it was shown in \cite{Katz:2011qp} that it is possible at least at leading order in the $SU(5)$ divisor, which we denote by\footnote{Since the $U(1)$s are global objects we require the fibration to take the Tate form at all orders in $w$.} 
\begin{equation}
W: w=0 \;.
\end{equation} 
Such Tate models which support an $SU(5)$ singularity on $w=0$ are given by the specialisation of the $a_i$ to the form \cite{Bershadsky:1996nh}
\begin{equation}
a_1 = {\rm generic} \;,\;\;a_2 = a_{2,1} w \;,\;\;a_3 = a_{3,2} w^2 \;,\;\;a_4 = a_{4,3} w^3 \;,\;\;a_6 = a_{6,5} w^5 \;. \label{tatesu5}
\end{equation} 
Here the $a_{i,n}$ are functions of the base coordinates, including $w$, but which do not vanish at $w=0$ so that the assignment (\ref{tatesu5}) fixes the vanishing order of the $a_i$ at $w=0$.

The class of sections we are interested in is, in the Tate form (\ref{tatesu5}), defined by the equation\footnote{Note that in  \cite{Marsano:2010ix,Marsano:2011hv}  the section (\ref{tatesection}) was termed the Tate divisor and was conjectured to be the global extension of the spectral cover. We discuss the relation to the local spectral cover more in appendix \ref{App-SCC}.}
\begin{equation}
y^2 = x^3 \;. \label{tatesection}
\end{equation} 
For generic $a_{i,n}$ this defines the zero section $z=0$ only. However for special forms of the $a_{i,n}$ it will define a whole class of sections, and these additional ones correspond to $U(1)$ symmetries. 

To deduce the form of the $a_{i,n}$ it is useful to rewrite (\ref{tatesection}) in terms of the variable
\begin{equation}
t \equiv \frac{y}{x}  \label{tdef}
\end{equation} 
as
\begin{equation}
x = t^2  \label{xt2}.
\end{equation} 
Note that (\ref{xt2}) implies that in specifying the section an equation in $t$ is holomorphic and well behaved at $x=0$. Now using (\ref{tdef}) we see that $Y_4$ is given by the vanishing locus of the  Tate polynomial 
\begin{equation}
P_T = x^2(x-t^2) + x^2 t z a_1 + x^2 z^2 a_{2,1} w + t x z^3 a_{3,2} w^2 + x z^4 a_{4,3} w^3 + z^6 a_{6,5} w^5 \;
\end{equation} 
inside $X_5$.
The section is specified by (\ref{xt2}) on $Y_4$, i.e.\ by the vanishing of
\begin{equation}
X = 0 \quad  \cap \quad P_T = 0
\end{equation} 
inside $X_5$, where we defined
\begin{equation}
X \equiv t^2-x \;.
\end{equation} 
Note that 
\begin{equation}
P_T|_{X=0} = t^5 z a_1 + t^4 z^2 a_{2,1} w + t^3 z^3 a_{3,2} w^2 + t^2 z^4 a_{4,3} w^3 + z^6 a_{6,5} w^5 \; \label{tpartsec}
\end{equation} 
and for generic polynomials $a_{i,n}$ the only holomorphic solution is at $z=0$.

In this form the condition for existence of further sections of the type $X=0$ becomes obvious, namely $P_T|_{X=0}$ must factorise holomorphically such as to allow for extra holomorphic zeroes in addition to the universal solution $z=0$, i.e.\ 
\begin{eqnarray}
P_T|_{X=0} =  - z\prod_{i=1}^n Y_i
\end{eqnarray}
for some holomorphic polynomials $Y_i$. This in turn implies
\begin{equation} \label{Ptfact}
P_T =  X Q - z\prod_{i=1}^n Y_i
\end{equation} 
with $Q$ a holomorphic polynomial as well, and therefore $Y_4$ is given by the hypersurface
\begin{equation}
X Q  = z\prod_{i=1}^n Y_i \quad   \subset X_5\;. \label{tatesing}
\end{equation} 
Once the polynomials $a_{i,n}$ are restricted in such a way that (\ref{Ptfact}) holds, the 4-fold $Y_4$ exhibits $n$ obvious sections
\begin{equation}
X = 0 \quad \cap \quad Y_i=0, \quad i=1, \ldots ,n \label{sections}
\end{equation} 
in addition to the zero section at $z=0$.
The relevance of global factorisations of the spectral cover equation in heterotic models with extra $U(1)$ symmetries has been investigated in \cite{Choi:2012pr}.

The sections (\ref{sections}) are not all independent because their product generically will include a term proportional to $z^5$ which is absent from the Tate form (\ref{tateform}). Therefore there is one constraint on their coefficients for such a term to be absent. This constraint is the tracelessness constraint. If we think of the $SU(5)$ as emerging as the commutant of an $SU(5)_\perp$ inside an underlying $E_8$ as is traditionally done in the context of the local Higgs bundle picture, the tracelessness constraint ensures that the $U(1)$s are embedded into $SU(5)_\perp$ rather than $U(5)_\perp$. In terms of the points on the torus corresponding to the sections it implies that the sum of them gives back the zero section at the origin \cite{Friedman:1997yq}. Thus, an $n$-fold factorisation as in (\ref{Ptfact}) corresponds to $n-1$  independent extra sections. 

An important aspect of the sections (\ref{sections}) is that the $Y_i$ are generally not linear polynomials. This means that the equation (\ref{sections}) for a given $Y_i$ in fact defines a number of points corresponding to the multiple roots of the polynomial $Y_i=0$. We define the section as the torus sum of these points (see \cite{Morrison:2012ei,Choi:2012pr} for example on how the addition of points on the torus fibre is performed). The individual roots themselves still hold information though as extra matter states localise on loci where two of the roots degenerate. In the presence of extra non-Abelian gauge symmetry $G$ these states are singlets under $G$. We will show that when the two roots come from different $Y_i$ factors the associated singlets are charged under the $U(1)$s. It is also natural to expect that when the roots are in the same $Y_i$ factor the singlets are neutral under all $U(1)$s, though such states are more difficult to identify as they do not correspond to a singularity on the 
manifold. We will show that it is also possible to combine these possibilities with two pairs of roots degenerating from each factor leading to four degenerate roots and in this case doubly charged singlets localise. We discuss this in more detail in section \ref{sec:fibu1pq} for a particular example.
 
To understand how the extra sections give rise to a $U(1)$ symmetry, we note that as it stands, (\ref{Ptfact}) is singular - even away from the obvious $A_4$ singularity in the fibre over the $SU(5)$ divisor $w=0$.
This is because the equation is in so-called binomial form, whose importance in F-theory was stressed more recently in \cite{Esole:2011sm,Braun:2011zm}. 
The singularities of $P_T$ not owed to the $SU(5)$ gauge group arise at the intersection of
\begin{eqnarray} \label{locsing}
X=0 \quad \cap \quad Q = 0 \quad \cap \quad Y_i=0  \quad \cap \quad Y_j=0 \;,
\end{eqnarray}
which for each pair of $i,j$ describes a curve of singularities. Note that because $X=0$ is part of the singularity it is in the patch where the variable $t$ is holomorphic and well defined. However we should be careful when analysing singularities on the particular locus $x=0$ since they depend on a derivative analysis which does not hold generally on this locus: the manifold could still remain smooth and the apparent singularity due to the binomial form is misleading. The potential singular nature of the 4-fold at $x=0$ must therefore be checked by going back to the original Weierstra\ss{} formulation of the model with fibre coordinates $x,y,z$. As we will show this only affects a certain class of $SU(5)$ singlets which localise on curves that we discuss in section \ref{sec:fibu1pq}.

In particular for the case of a single $U(1)$, and therefore two splitting factors $Y_1$ and $Y_2$, we have a conifold singularity over the curve $X=Q=Y_1=Y_2=0$ \cite{Grimm:2010ez, Braun:2011zm}. In fact the fibre over this curve exhibits an $SU(2)$ singularity so that we will refer to this singular locus as the curve of $SU(2)$ singularities. These singularities must be resolved. We will denote the resolved 4-fold by $\hat Y_4$.\footnote{We do not distinguish in notation between the 4-fold where only the $SU(2)$ singularities are resolved or where also the $SU(5)$ singularity over the divisor $W$ are resolved.} The resolution introduces new divisor classes $S_i$ in $H^{1,1}(\hat Y_4)$. These $S_i$ are then related to elements $\tw_i \in H^{1,1}(\hat Y_4)$ such that expansion of the M-theory 3-form $C_3$ as 
\begin{equation}
C_3 = A_i \wedge \tw_i + \ldots
\end{equation} 
gives rise to gauge potentials $A_i$ of the Abelian symmetry $A_i$. We will determine the relation between $S_i$ and $\tw_i$ in detail in section \ref{sec:G4-41}.
 
This approach allows for a systematic construction of extra $U(1)$ symmetries for Tate models by classifying all possible factorisations of the $P_T|_{X=0}$ of the form (\ref{Ptfact}). Since $P_T|_{X=0}$ is a polynomial of degree $5$ in $t$, this amounts to making a general ansatz for the degree $n_i$ polynomials $Y_i$ with $\sum_i n_i =5$, subject to extra constraints such that $P_T|_{X=0} = - z\prod_i Y_i$. 
For example, if we are interested in one extra section, there are two inequivalent classes of factorisations because the degrees of $Y_1$ and $Y_2$ can be $(n_1,n_2) = (1,4)$ or $(2,3)$. We will give the explicit form of the factors $Y_1$, $Y_2$ and $Q$ in section \ref{sec_Split-Tate}. 

Note that a specific  global $SU(5)$ Tate model with one extra $U(1)$ was introduced as the $U(1)$ restricted Tate model in \cite{Grimm:2010ez}, and \cite{Braun:2011zm} showed that this model can be brought in the form (\ref{Ptfact}) with $n=2$. We elaborate further on the relation of the factorised Tate models to the $U(1)$ restricted Tate model approach in appendix \ref{sec:reltate}.

The class of global Tate models has a well-defined local limit $w \rightarrow 0$, in which it flows to the so-called spectral cover or Higgs bundle construction of local models \cite{Donagi:2009ra}. We review this limit in appendix  \ref{App-SCC}. Correspondingly, our factorised Tate models (\ref{Ptfact}) precisely flow to what is called split spectral cover models in the local F-theory literature \cite{Donagi:2009ra,Marsano:2009gv,Marsano:2009wr}. It is therefore clear that the constraints on the coefficients $a_{i,n}$ are identical to the constraints on the sections on $W$ which define the split spectral covers.
It is important to stress, though, that the existence of a $U(1)$ symmetry and the associated $U(1)$ fluxes can never be determined in a satisfactory manner by focusing only on the local limit. Concretely the factorised Tate model constrains also higher order terms in $w$ which do not feature in the spectral cover limit. The $U(1)$ symmetry is sensitive to the full global details of the compactification \cite{Hayashi:2010zp,Grimm:2010ez}. This in particular requires a full resolution of the binomial singularities (\ref{locsing}) to determine the resolved version of the extra sections \cite{Grimm:2010ez,Braun:2011zm,Krause:2011xj,Grimm:2011fx}. The factorised Tate model (\ref{Ptfact}) can be viewed as the correct global extension of the local split spectral cover models. 

\subsection{Resolving the \texorpdfstring{$SU(5)$}{SU(5)} singularity}

The discussion just presented was phrased in the limit where $Y_4$ exhibits an $SU(5)$ singularity in the fibre over $w=0$. In order to fully analyse the model, however,  we are interested in understanding the sections after resolving the $SU(5)$ singularity. This resolution process  has been studied with  different techniques in the recent F-theory $SU(5)$ GUT literature in a number of papers \cite{Blumenhagen:2009yv,Esole:2011sm,Marsano:2011hv,Krause:2011xj,Grimm:2011fx,Krause:2012yh} (see  \cite{Chen:2010ts} 
 for other gauge groups) and we use the process described in \cite{Krause:2011xj,Krause:2012yh}. The resolution is achieved through a sequence of 4 blow-ups. This introduces 4 resolution divisors $E_i: e_i=0, \, i=1, \ldots,4$ and amounts to the replacement
\begin{equation}
x \rightarrow x e_1 e_2^2 e_3^2 e_4 \;,\;\; y \rightarrow y e_1 e_2^2 e_3^3 e_4^2 \;,\;\; w \rightarrow e_0 e_1 e_2 e_3 e_4 \;. \label{restrans}
\end{equation} 
Accordingly the Tate polynomial reads
\begin{eqnarray}
P_T = e_1^2 e_2^4 e_3^5 e_4^3 & [x^3 e_1 e_2^2 e_3 - y^2 e_4 e_3 + a_1 x y z + a_{2,1} x^2 z^2 e_0 e_1 e_2 + a_{3,2} y z^3 e_0^2  e_1 e_4 \nn  \\
& + a_{4,3} x z^4 e_0^3 e_1^2 e_2 e_4 +  a_{6,5} z^6 e_0^5 e_1^3 e_4^2 e_2 ] \;.\label{restate}
\end{eqnarray}
The proper transform $\hat P_T$ is obtained by dividing by the overall factor and describes the resolved Calabi-Yau 4-fold $\hat Y_4$ as the hypersurface
\begin{equation}
\hat P_T = x^3 e_1 e_2^2 e_3 - y^2 e_4 e_3 + a_1 x y z + a_{2,1} x^2 z^2 e_0 e_1 e_2 +a_{3,2} y z^3 e_0^2  e_1 e_4 + a_{4,3} x z^4 e_0^3 e_1^2 e_2 e_4 + a_{6,5} z^6 e_0^5 e_1^3 e_4^2 e_2 \;\label{restatept}
\end{equation} 
inside an ambient 5-fold $\hat X_5$.
This ambient space $\hat X_5$ of the resolution is subject to a rich Stanley-Reisner ideal given by \cite{Krause:2011xj,Krause:2012yh}
\begin{eqnarray}
\{x y z,\, x y e_0,\, x e_0 e_3,\, x e_1 e_3,\, x e_4,\, y e_0 e_3,\, y e_1,\, y e_2,\, z e_1 e_4,\, z e_2 e_4,\, z e_3,\, e_0 e_2\}  \label{srimain}
\end{eqnarray}
and one possible choice from the combinations
\begin{equation}\label{srideal_optional}
 \left\{
  \begin{aligned}
   & y e_0 \\ & z e_4
  \end{aligned}
 \right\}
 \otimes
 \left\{
  \begin{aligned}
   & x e_0,\, x e_1 \\ & x e_0,\, z e_2 \\ & z e_1,\, z e_2
  \end{aligned}
 \right\}
 \otimes
 \left\{
  \begin{aligned}
   & e_0 e_3,\, e_1 e_3 \\ & e_0 e_3,\, e_2 e_4 \\ & e_1 e_4,\, e_2 e_4
  \end{aligned}
 \right\}.
\end{equation}
The different choices correspond to different triangulations. For definiteness we will work in the sequel with one particular triangulation corresponding to the choice of elements
\begin{equation}
\{x y z, \, z e_i|_{i=1, \ldots,4},\, x y e_0,\, x e_0 e_3,\, x e_1 e_3,\, x e_4,\, y e_0 e_3,\, y e_1,\, y e_2,\, z e_1 e_4,\, z e_2 e_4,\, e_0 e_2, e_4 e_1,\, e_4 e_2 \} \;. \label{sritriang}
\end{equation} 
Note, however, that the specific form of the resolved fibre may dependent on the concrete triangulation under consideration.

We now wish to apply the same logic  as in section \ref{sec_factorisation} to describe $U(1)$s after the $SU(5)$ resolution. The first thing to specify is the class of sections analogous to (\ref{tatesection}). We take this to be
\begin{equation}
y^2 e_4 = x^3 e_1 e_2^2 \; \label{tatesecres}
\end{equation} 
in view of the quantities appearing in (\ref{restatept}) after dividing by a factor of $e_3$.
We define $t$ as in (\ref{tdef}) but with $x$ and $y$ the  coordinates appearing in (\ref{restatept}). Note that because the coordinate transformation (\ref{restrans}) acting on $t$ is holomorphic in the $e_i$ the potential subtlety discussed in the previous section remains only on the locus $x=0$. Suppose the Tate model prior to $SU(5)$ resolution takes the factorised form (\ref{Ptfact}). Since all we have done to arrive at (\ref{restate}) is to transform coordinates as (\ref{restrans}), the resulting $P_T$ as given in (\ref{restate}) is guaranteed to factorise on the locus $X=0$, where now
\begin{equation}
X = t^2 e_4 - x e_1 e_2^2 \;. 
\end{equation} 
However, what is not guaranteed is that the proper transform $\hat P_T$ given in (\ref{restatept}) also factorises into \emph{holomorphic} components since we have divided out by the prefactor in (\ref{restate}). Indeed it does not. This can be checked on a case-by-case basis as demonstrated in the following sections. However the meromorphicity arises purely from the resolution divisors $e_1$ and $e_2$. With the choice of triangulation (\ref{sritriang}) it is simple to check that these divisors do not intersect the section and therefore the singularity because the Stanley-Reisner ideals forbids the intersection of (\ref{tatesecres}) with $e_1=0$ and $e_2=0$. In fact also $e_4$ does not intersect the section and so only $e_0$ and $e_3$ are relevant.

The result that $e_1$, $e_2$ and $e_4$ do not intersect the section implies that in order to resolve the binomial singularity (\ref{locsing}) we can work in a patch where we set $e_1=e_2=e_4=1$. In this patch the resolved Tate form does split holomorphically over the section (\ref{tatesecres}) and can be again written as (\ref{tatesing}) with holomorphic $Q$ and the $Y_i$. Therefore, in this patch, we can resolve the additional singularity and account for the $U(1)$s. This will be worked out for the individual factorisations in the next section.

Let us stress that in section \ref{sec:relmorpa} we will provide a rather different resolution of the factorised Tate models based on a $\mathbb P_{1,1,2}[4]$-fibration that had appeared before in \cite{Morrison:2012ei}. In this approach we will not need to work patchwise, which is more gratifying from a formal perspective. However, since the actual structure of the factorised Tate models is more evident and intuitive in the current framework we find it useful to present the analysis of the matter spectrum etc. in this fashion in sections \ref{sec:fibstruc-gen} and \ref{sec:fibu1pq}.

\subsection{Factorised Tate Models} \label{sec_Split-Tate}

Having outlined the general approach and formalism we can tackle the specific factorisations as given in table \ref{tab:breake8u1s}.
In this section we work out the explicit form of the equations for the $4-1$ and the $3-2$ factorisations. The other cases are presented in appendix \ref{sec:morfactate}.
\begin{table}
\centering
\begin{tabular}{|l|c|}
\hline
Factorisation Pattern & Number of $U(1)$s \\
\hline
 $Y^{(1)}_1 Y^{(4)}_2 $& 1 \\
\hline
$ Y^{(2)}_1 Y^{(3)}_2 $ & 1 \\
\hline
 $Y^{(1)}_1 Y^{(1)}_2  Y^{(3)}_1$  & 2 \\
\hline
$Y^{(1)}_1 Y^{(2)}_2  Y^{(2)}_3  $  & 2 \\
\hline
$Y^{(1)}_1 Y^{(1)}_2  Y^{(1)}_3  Y^{(2)}_4$    & 3 \\
\hline
$Y^{(1)}_1 Y^{(1)}_2  Y^{(1)}_3  Y^{(1)}_4 Y^{(1)}_5 $  & 4 \\
\hline
\end{tabular}
\caption{Possible factorisation patterns of $P_T|_{X=0}$. The superscripts denote the degree in $t$.}
\label{tab:breake8u1s}
\end{table}

\subsubsection{\texorpdfstring{$4-1$}{4-1} Factorisation}
\label{sec:41split}

The $4-1$ factorisation corresponds to writing $\hat P_T|_{X=0} =  - z Y_1 Y_2$ with $Y_1$ and $Y_2$ polynomials of respective degrees 1 and 4 in $t$. Performing the resolution (\ref{restrans}) on the general factorised form gives, after the proper transform, 
\begin{eqnarray}
Y_1 = c_1 t + c_0 e_0 z, \qquad Y_2 = e_4^2 \left(t^4 d_4 + t^3 e_0 z d_3 + t^2 e_0^2 z^2 d_2 + d_1 t e_0^3 z^3 + d_0 e_0^4 z^4 \right).
\end{eqnarray}
Here we have set $e_1 = e_2 =1$ because, as discussed in the previous section, these two resolution divisors do never intersect the extra section.
Comparing the above ansatz with $\hat P_T|_{X=0}$ reveals that the polynomials $c_i$ and $d_i$ are subject to the tracelessness constraint 
\begin{eqnarray} \label{Constr1}
c_1 d_0 + c_0 d_1=0
\end{eqnarray}
because there is no term of order $t$ in $\hat P_T|_{X=0}$. As discussed this is a consequence of the fact that the Tate model  has no $z^5$ term.

As mentioned before and discussed in greater detail in appendix \ref{App-SCC}, the factorised Tate model asymptotes, in the local limit $w \rightarrow 0$, to the split spectral cover models. Indeed the factorisation structure and in particular the constraint (\ref{Constr1}) are as for the $U(1)_X$ spectral cover worked out in \cite{Marsano:2009gv}. The solution to this constraint can be written as
\begin{equation} \label{d0d1X}
d_0 = \alpha c_0 \;,\;\; d_1 = -\alpha c_1 \;
\end{equation} 
with $\alpha$ some polynomial on $B$ of appropriate degree.
Note that we must impose that $c_0$ and $c_1$ should not vanish simultaneously in order not to induce non-Kodaira singular fibres because at this locus all the $a_{n,i}$ vanish. 

Given $Y_1$ and $Y_2$ we can now explicitly evaluate also the polynomial $Q$ and arrive at the following parametrisation of the $4-1$ factorised Tate model,
\begin{eqnarray}
Y_1 &=& t c_1 + c_0 u \;, \nn \\
Y_2 &=& e_4^2 \left(t^4 d_4 + t^3 u d_3 + t^2 u^2 d_2 - \alpha c_1 t u^3 + \alpha c_0 u^4 \right) \nn \\
X &=& t^2 e_4 - x \;, \label{ptf41} \\
Q &=& e_3 x^2 + c_1 d_4 e_4 t^3 z + c_1 d_3 e_4 t^2 u z + c_0 d_4 e_4 t^2 u z + c_1 d_2 e_4 t u^2 z + 
 c_0 d_3 e_4 t u^2 z - \alpha e_4 c_1^2 u^3 z \nn \\ & &+ c_0 d_2 e_4 u^3 z + c_1 d_4 t x z + 
 c_1 d_3 u x z + c_0 d_4 u x z\; \nn 
\end{eqnarray}
with $u=e_0 z$. The case where the $SU(5)$ is unresolved is reached simply by setting $e_3=e_4=1$ and $e_0=w$.%

Finally let us briefly describe the resolution of the binomial singularity $X=Q=Y_1=Y_2=0$. We stress again that this singularity lies entirely in the patch $e_1=e_2=1$.
Such  type of binomial singularities has been introduced recently in \cite{Esole:2011sm} in the context of $SU(5)$ models without extra $U(1)$s and in \cite{Braun:2011zm}, which has brought the $U(1)$ restricted Tate model of \cite{Grimm:2010ez} into binomial form. The small resolution proceeds by replacing the singularity in the fibre over  the curve $X=Q=Y_1=Y_2=0$ by a $\mathbb P^1$ parametrised by homogeneous coordinates $[\lambda_1,\lambda_2]$. This is achieved by describing $\hat  Y_4$ in the given patch as the complete intersection
\begin{equation} \label{compint}
 Y_1 \lambda_2 = Q \lambda_1\quad  \cap \quad Y_2 \lambda_1 = X \lambda_2  
\end{equation} 
inside a 6-fold $\hat X_6$. 
Away from $X=Q=Y_1=Y_2=0$, the extra section is given by the locus
\begin{eqnarray}
\lambda_1=0 \quad \cap \quad X=0 \quad \cap \quad Y_1 = 0 \label{y1secsim}
\end{eqnarray}
inside $\hat X_6$ as follows by plugging $X=Y_1=0$ into (\ref{compint}). At $X=Q=Y_1=Y_2=0$, on the other hand,  $[\lambda_1,\lambda_2]$ are unconstrained and therefore the section wraps the entire resolution $\mathbb P^1$ as in  \cite{Grimm:2010ez,Braun:2011zm}. This behaviour will be discussed in greater detail in section \ref{sec:fibstruc-gen}.
\subsubsection{\texorpdfstring{$3-2$}{3-2} Factorisation}
\label{sec:32split}

The $3-2$ factorised Tate model is based on the ansatz
\begin{equation}
Y_1 = c_2 t^2 + c_1 t e_0 z + c_0 e_0^2 z^2, \qquad Y_2 = e_4^2 (d_3 t^3 + d_2 e_0 t^2 z  + d_1 t z^2 + d_0 e_0^3 z^3)      
\end{equation} 
subject to the constraint
\begin{equation}
c_1 d_0 + c_0 d_1
\end{equation} 
from $a_5=0$ in the Tate polynomial.
As in the local split spectral cover version a way to solve the tracelessness constraint is to write \cite{Dolan:2011iu}\footnote{Note that it seems we are writing 4 parameters in terms of 4 other parameters while solving a constraint $a_5=0$, which is not possible. Indeed there are only 3 independent parameters in the ansatz (\ref{31tra}) since one can write $c_1=c_0 \left(\frac{\delta}{\beta}\right)$ and $d_1=d_0\left(\frac{\delta}{\beta}\right)$. However the important point noted in \cite{Dolan:2011iu} is that taking the solution (\ref{31tra}) allows for additional freedom in distributing the possible globally trivial components of the matter curves and so can be important when considering the restriction of hypercharge flux to matter curves.}
\begin{eqnarray}
c_0 = \alpha \beta \;,\;\;\; c_1 = \alpha \delta \;, \;\;\; 
d_0 = \gamma \beta \;,\;\;\; d_1 = -\gamma \delta \; \label{31tra}
\end{eqnarray}
with $\alpha$, $\beta$, $\gamma$ and $\delta$ arbitrary polynomials of appropriate degrees. In order to forbid non-Kodaira singularities one must impose that the following intersections should be empty,
\begin{equation} \label{3-2-Kodaira}
c_2 \cdot \alpha \;,\; c_2 \cdot \beta \cdot \delta \;,\; d_2 \cdot d_3 \cdot \gamma \;.
\end{equation} 
It is worth noting that the constraints will expand if some of the factors are set to zero identically over the full 4-fold.

With this information one can again compute $Q$ and arrive, in the patch $e_1=e_2=1$, at the binomial form
\begin{eqnarray}
Y_1 &=& c_2 t^2 + \alpha \delta e_0 t z + \alpha \beta e_0^2 z^2 \;, \nn \\
Y_2 &=& e_4^2 (d_3 t^3 + d_2 e_0 t^2 z - \delta e_0^2 \gamma t z^2 + \beta e_0^3 \gamma z^3) \;, \nn \\
X &=& t^2 e_4 - x \;,  \\
Q &=& e_3 x^2 + c_2 d_3 e_4 t^3 z + c_2 d_3 t x z + c_2 d_2 e_0 e_4 t^2 z^2 + 
 \alpha d_3 \delta e_0 e_4 t^2 z^2 + c_2 d_2 e_0 x z^2  \nn \\
& & + \alpha d_3 \delta e_0 x z^2 + 
 \alpha \beta d_3 e_0^2 e_4 t z^3 + \alpha d_2 \delta e_0^2 e_4 t z^3 - 
 c_2 \delta e_0^2 e_4 \gamma t z^3 + \alpha \beta d_2 e_0^3 e_4 z^4  \nn \\
&& + \beta c_2 e_0^3 e_4 \gamma z^4 - \alpha \delta^2 e_0^3 e_4 \gamma z^4 \;. \nn
\end{eqnarray}

\section{ Fibre Structure and charges in the \texorpdfstring{$U(1)_X$}{U(1)X} model}
\label{sec:fibstruc-gen}

We now analyse the fibre structure of our F-theory compactification, starting, in this section, with the $4-1$ factorisation. The associated Abelian gauge symmetry is often referred to as $U(1)_X$ in the model building literature.
Over generic points on the $SU(5)$ divisor $W: w=0$ in the base $B$, the fibre can be described in terms of the hypersurface equation (\ref{restatept}) within $\hat X_5$. 
The resolution of the $SU(2)$ singular locus is described by the complete intersection (\ref{compint}) within $\hat X_6$. The Yukawa points on $W$ where $SU(5)$ matter couples to the singlets localised along the $SU(2)$ curve can also be treated in this approach as we will see.  In total this gives us access to the fibre structure over the entire Calabi-Yau $ \hat Y_4$.

\subsection{Structure of the matter surfaces} \label{sec:fibstruc}

We begin with the fibre structure over $W$. 
In analysing the resolution $\mathbb P^1$s over $W$ and over the various matter curves we follow the procedure described in \cite{Krause:2011xj} (see also \cite{Esole:2011sm,Marsano:2011hv,Grimm:2011fx,Krause:2012yh,Chen:2010ts}).

As usual the fibre over generic points on the $SU(5)$ surface $W: w=0$ in the base $B$ is given by a tree of $\mathbb P^1$s intersecting like the affine Dynkin diagram of $SU(5)$. These $\mathbb P^1_i, i=0, \ldots, 4$ are the fibres of the resolution divisors $E_i: e_i=0$ and can be described as the complete intersection
\begin{eqnarray} \label{P1W}
\pl{i}: \quad  e_i = 0 \quad \cap \quad \hat P_T=0  \quad \cap \quad D_a=0  \quad \cap \quad D_b=0   \quad \subset \hat X_5
\end{eqnarray}
with $D_a, D_b$ denoting the pullback of two base divisors that intersect $W$ exactly once.
The intersection of these divisors is such that
\begin{eqnarray} \label{Cij}
\int_{\hat Y_4} E_i \wedge E_j \wedge D_a \wedge D_b = C_{ij} \int_B W \wedge D_a \wedge D_b
\end{eqnarray}
 with $C_{ij}$ the Cartan matrix of $SU(5)$ in conventions where the diagonal has entries $-2$.

Over the matter curves on $W$ some of these $\mathbb P^1$s split and assemble into the affine Dynkin diagram of higher rank groups.

\subsubsection*{10-matter curves} 

We first turn to the intersection curve of the Tate polynomial $a_1=0$ with $W$ in the base, which, in an $SU(5)$ Tate model, corresponds to the ${\bf 10}$ matter curve.
Since in the $U(1)_X$ model $a_1 = c_1 \, d_4$ there are now two ${\bf 10}$ curves 
\begin{eqnarray} \label{U1X10curve}
C_{{\bf 10}^{(1)}}:  \quad  d_4 = 0 \quad \cap \quad w=0, \qquad \quad C_{{\bf 10}^{(2)}}:  \quad c_1 = 0 \quad \cap  \quad w=0.
\end{eqnarray}
As discussed in greater detail in appendix \ref{App-SCC} the structure of the $SU(5)$ charged matter curves coincides with the corresponding split spectral cover model \cite{Marsano:2009gv} to which our construction flows near the $SU(5)$ divisor. Note that  $C_{{\bf 10}^{(2)}}$ did not appear in the $U(1)$ restricted Tate model of \cite{Krause:2011xj}, where $c_1$ was set to $1$, see appendix \ref{sec:reltate}.

The  ${\mathbb P}^1$-structure over both ${\bf 10}$-curves turns out to be very similar. To describe the fibre we must specialise, say, $D_b=0$ in (\ref{P1W}) to $d_4=0$ or $c_1=0$, respectively. As a consequence the polynomial $\hat{P}_T$ will factorise for certain $\pl{i}$. Such a factorisation indicates a splitting of $\mathbb P^1_i$s over the matter curve.

Concretely over $C_{{\bf 10}^{(1)}}$ this procedure yields the following equations (omitting for brevity the universal piece $d_4=0 \cap D_a=0$) and corresponding $\mathbb P^1$-splits, 
\begin{eqnarray} \label{P1-hyperA}
e_0 = 0 \quad \cap \quad    e_3 \,  (-e_1 + e_4) = 0 &\longleftrightarrow&    \pl{0} \rightarrow \mathbb P^1_{03} \cup \mathbb P^1_{0A}, \nn \\
e_1=0  \quad  \cap \quad   e_3 =0  &\longleftrightarrow&    \pl{1} \rightarrow   \mathbb P^1_{13},  \nn \\
e_2=0  \quad  \cap \quad   e_3 - (c_1 d_2 + c_0 d_3)\, e_1 \, z^3 =0 &\longleftrightarrow&    \pl{2} \rightarrow   \mathbb P^1_{2B_1},   \\
e_3=0  \quad  \cap \quad e_0 \, e_1 \, C = 0 &\longleftrightarrow&    \pl{3} \rightarrow   \mathbb P^1_{03} \cup   \mathbb P^1_{13} \cup   \mathbb P^1_{3C_1}, \nn  \\
e_4=0  \quad  \cap \quad  e_3 + c_1 \, d_3 \, e_0 \, z^2  =0 &\longleftrightarrow&    \pl{4} \rightarrow \mathbb P^1_{4D_1}, \nn
\end{eqnarray}
where we have exploited the Stanley-Reisner ideal to set as many coordinates to one as possible.\footnote {The polynomial $C_1$ takes the form $C_1= c_1 e_2 x (d_1 e_0^2 e_1 e_4 + d_3 x ) + c_1 d_2 e_0 e_4 y + c_0 e_0 e_4 (e_0 e_1 e_2 (d_0 e_0^2 e_1 e_4 + d_2 x ) + d_3 y)$.}  %
  A factorisation of the above defining equations indicates a splitting of $\mathbb P^1$s over the matter curve into the indicated $\mathbb P^1$s. Note that $\mathbb P^1_{13}$ and   $\mathbb P^1_{03}$ appear with multiplicity two. 
It is now easy to compute the intersection structure of these six $\mathbb{P}^1$s by counting simultaneous solutions to these equations within $\hat X_5$.   For instance, since $e_2 e_4$ is in the Stanley-Reisner ideal, $\pl{4D_1}$ and $\pl{2 B_1}$ do not intersect. On the other hand, $\pl{03} \cap \pl{4D_1} =1$ because $D_1|_{e_3=0=e_1}$ vanishes identically so that this intersection is described by the transverse intersection of five polynomials $e_0=e_3=e_4=d_4=D_a=0$ within $\hat X_5$.
  In this fashion one establishes that the six $\mathbb P^1$s  intersect like the nodes of the affine Dynkin diagram of $SO(10)$ as required in the theory of Kodaira fibres.

Over  $C_{{\bf 10}^{(2)}}$, the exact form the of the defining equations differs slightly, but the $\mathbb P^1s$ split in an analogous manner into
$\mathbb P^1_{0A}$, $\mathbb P^1_{23}$, $\mathbb P^1_{2B_2}$, $\mathbb P^1_{3C_2}$, $\mathbb P^1_{4B_2}$,  each with multiplicity one, and $\mathbb P^1_{13}$ and   $\mathbb P^1_{03}$ each with multiplicity two. The intersection structure is again as in the affine Dynkin diagram of $SO(10)$.

To identify the combinations of $\mathbb P^1s$ corresponding to the ${\bf 10}$ representation one must compute the Cartan charges of the $\mathbb P^1s$  and compare these to the ${\bf 10}$ weights. 
We observe that the structure of the matter surfaces is identical to the ${\bf 10}$ curve in the $U(1)$ restricted Tate model as analysed in  \cite{Krause:2011xj} for the analogous choice of Stanley-Reisner ideal. 
Therefore we can refer to \cite{Krause:2011xj}, section 3.3  for the computation of the $SU(5)$ Cartan charges of the above $\mathbb P^1s$ and to tables A.18 and A.19 for  the resulting identification of suitable combinations of $\mathbb P^1s$ with the weight vectors of the ${\bf 10}$ representation of $SU(5)$.  
For convenience of the reader we recall this procedure for the $SU(5)$ Cartan charges of $\mathbb P^1_{03}$ over $C_{{\bf 10}^{(1)}}$. 
The gauge potential associated with the Cartan $U(1)_i \subset SU(5)$ arises by expanding the M-theory 3-form as $C_3 = A_i \wedge E_i + \ldots$, where $E_i$ denotes the 2-form dual to the resolution divisor $e_i=0$.
Therefore the charge under the generator of $U(1)_i \subset SU(5)$ is given by the integral $\int_{\mathbb P^1_{03}} E_i$, $i=1, \ldots,4$. This can be computed as the intersection
\begin{eqnarray}
e_i = 0 \quad \cap \quad e_0 = 0 \quad \cap \quad e_3=0 \quad \cap \quad d_4=0 \quad \cap \quad D_a =0\quad \subset \hat X_5.
\end{eqnarray}
For $i=1$ and $i=4$ this is the transverse intersection of five degree-one polynomials inside $\hat X_5$, which have one intersection point. For $i=2$, on the other hand, this vanishes because $e_0 e_2$ is in the Stanley-Reisner ideal. Finally for $i=3$ we do not encounter an effective intersection must therefore use the following trick: We first note that the integral $\int_{\mathbb P^1_{0A}} E_3 =1$, and that the generic intersection of $e_3=0$ with $\mathbb P^1_0$ over a generic point on the $SU(5)$ divisor $W: w=0$ in the base vanishes because of (\ref{Cij}). Since ${\mathbb P}^1_0$ splits into $\pl{03}$ and $\pl{0A}$, this implies $\int_{\mathbb P^1_{03}} E_3 = -1$.
Therefore the $U(1)_i$ charges of $\pl{03}$ are $[1,0,-1,1]$, corresponding to the weight  $\mu_{\bf 10} - \alpha_2 - \alpha_3$ of the ${\bf 10}$ representation of $SU(5)$.

The Cartan charges over the second ${\bf 10}$ curve work out in exactly the same manner. 
For convenience the $\pl{}$-combination for the various states of the $\mathbf{10}$-representation is summarised in the following table (valid for both {\bf 10} matter curves):
\begin{equation}\label{10_1:p1_combination}
 \begin{array}{c|c}
  \textrm{Weight}                                           &     \pl{i}-{\rm combination}    \\
  \hline
  \hline
  \mu_{10}                                                                                     &    2 \pl{03} + \pl{0A}+ \pl{13} + \pl{4D_i}          \\
  \mu_{10} - \alpha_2                                                                  &    2 \pl{03} +  \pl{0A} + \pl{13} + \pl{2B_i}               + \pl{4D_i}        \\
  \mu_{10} - \alpha_1 - \alpha_2                                               &    2 \pl{03} +  \pl{0A} + 2\pl{13} + \pl{2B_i} +  + \pl{4D_i}        \\
  \mu_{10} - \alpha_2 - \alpha_3                                               &   \pl{03}             \\
  \mu_{10} - \alpha_1 - \alpha_2 - \alpha_3                            &  \pl{03}  + \pl{13}            \\
  \mu_{10} - \alpha_2 - \alpha_3 - \alpha_4                            &   \pl{03}  + \pl{4D_i}         \\
  \mu_{10} - \alpha_1 - 2\alpha_2 - \alpha_3                          &  \pl{03}  + \pl{13} + \pl{2B_i}        \\
  \mu_{10} - \alpha_1 - \alpha_2 - \alpha_3 - \alpha_4         &  \pl{03} + \pl{13} +  \pl{4D_i}       \\
  \mu_{10} - \alpha_1 - 2\alpha_2 - \alpha_3 - \alpha_4       & \pl{03}  + \pl{13} + \pl{2B_i}  + \pl{4D_i}     \\
  \mu_{10} - \alpha_1 - 2\alpha_2 - 2\alpha_3 - \alpha_4     & 2 \pl{03}  +  \pl{2B_i} +   \pl{3C_i} + \pl{4D_i}     \\
 \end{array}
\end{equation}

To see the difference between both ${\bf 10}$ curves we must investigate the intersection pattern of the $\mathbb P^1s$ with the extra section $S$. 
Since the resolution divisors $e_1$, $e_2$ and $e_4$ do not intersect the section, the only possible intersections occur for  $\mathbb P^1_{03}$ as well as for $\mathbb P^1_{3C_1}$ and $\mathbb P^1_{3C_2}$ (over $d_4=0$ or $c_1=0$, respectively). 
 Therefore it is sufficient to carry out the analysis of the intersection pattern with $S$ inside the complete intersection $\hat X_6$.
Recall that at generic points away from the $SU(2)$ singularities the resolved section $S$ is given by the locus 
\begin{eqnarray}
\lambda_1=0 \quad \cap \quad X=0 \quad \cap \quad Y_1 = 0
\end{eqnarray}
inside the complete intersection (\ref{compint}).
The intersection number between one of the above $\mathbb P^1s$ and the extra section $S$  is counted by the number of generic simultaneous solutions of the defining equations within $\hat X_6$. A priori these are seven constraints within the ambient 6-fold and thus have no common solution. However, it can happen that not all of these are independent. If we end up with precisely 6 mutually non-exclusive independent constraints, the intersection number is non-zero.

To simplify the expressions we will set $e_1=1, e_2=1, e_4=1$  as these are non-vanishing in the complete intersection patch and also $z=1$ because $z e_3$ is in the Stanley-Reisner ideal.
We start with the fibre over $C_{{\bf 10}^{(1)}}$ corresponding to $d_4=0$.
Along $\mathbb P^1_{03}$, the two constraints $Y_1=0$ and $X=0$ appearing in $S$ 
 evaluate to
\begin{eqnarray}
c_1 t = 0 \quad \cap \quad t ^2 - x =0.
\end{eqnarray}
The only solution over generic points on $d_4=0$ is $t=0=x$ and thus $x=y=0$, but $x y e_0$ is in the Stanley-Reisner ideal.
Thus $\mathbb P^1_{03}$ does not intersect the section in the fibre over $d_4=0$.

Concerning the intersection of $\mathbb P^1_{3C_1}$ with the section we note that the constraint $C_1=0$ is automatically fulfilled once we set $e_3=0$ and $Y_1 = X =0$.  Therefore we end up with the six independent constraints 
\begin{eqnarray}
x=t^2 \quad \cap \quad  c_1 t + c_0 e_0 =0 \quad \cap \quad e_3=0 \quad \cap \quad D_a=0 \quad \cap \quad \lambda_1=0 \quad  \cap \quad d_4=0.
\end{eqnarray}
It is important to note that in the present case $Y_1= c_1 t + c_0 e_0$ is of degree one in $t$. Therefore this system of equations has precisely one solution and thus $S \cap \mathbb P^1_{3C_1} =1$.

Over the second ${\bf 10}$ matter curve $C_{{\bf 10}^{(2)}}$, corresponding to $c_1=0$, the situation is reversed: 
For $e_0 = 0 = e_3 = c_1$, $Y_1$ vanishes automatically. Thus the intersection of $\mathbb P^1_{03}$ with $S$ is given by the single generic intersection of the six polynomials
\begin{eqnarray}
e_0 = 0  \quad \cap \quad  e_3 =0   \quad \cap \quad c_1=0 \quad \cap \quad D_a =0 \quad \cap \quad  \lambda_1 = 0 \quad \cap \quad  {x=t^2}
\end{eqnarray}
within the ambient 6-fold $\hat X_6$.
By contrast, now $\mathbb P^1_{3C_2}$ has no generic intersection with $S$ because on this locus $Y_1$ evaluates to $c_0 e_0$, and the intersection with $e_0=0$ had been accounted for already in $\mathbb P^1_{03}$.

To conclude, the difference between the two {\bf 10} curves is the following intersection pattern with the extra section $S$:
\begin{eqnarray}\label{ScapP}
&C_{{\bf 10}^{(1)}}:&   S  \cap  \mathbb P^1_{03} = 0, \qquad  S  \cap  \mathbb P^1_{13} = 0, \qquad S  \cap  \mathbb P^1_{3C_1} = 1, \nn \\
&C_{{\bf 10}^{(2)}}:&   S \cap  \mathbb P^1_{03} = 1,  \qquad  S   \cap  \mathbb P^1_{13} = 0, \qquad    S  \cap  \mathbb P^1_{3C_2} = 0.
\end{eqnarray}
This difference will be crucial when it comes to computing the $U(1)_X$ charges of the states.

\subsubsection*{5-matter curves} 

A similar analysis is easily carried out for the $\bf 5$ matter curves. For the $4 -1$ factorisation, the ${\bf 5}$ curve $P=0 \cap w=0$ in the base $B$ splits in the following way:
\begin{equation}
 P=a_1^2 a_{6,5} - a_1 a_{3,2} a_{4,3} + a_{2,1} a_{3,2}^2\quad\rightarrow\quad (d_3^2 c_0 + d_2 d_3 c_1 - d_1 d_4 c_1) (d_4 c_0^2 + d_3 c_0 c_1 + d_2 c_1^2)=: P_1 P_2\,,
\end{equation}
where we used the tracelessness constraint $d_1 c_0+c_1 d_0=0$. Like in the $U(1)$-restricted case of~\cite{Krause:2011xj}, over both {\bf 5}-curves 
\begin{eqnarray}
C_{{\bf 5}^{(1)}}: P_1 = 0 \quad \cap \quad w=0, \qquad C_{{\bf 5}^{(2)}}: P_2 = 0 \quad \cap \quad w=0
\end{eqnarray}
one observes a splitting of $\mathbb P^1_3$ into two $\mathbb P^1$'s,
\begin{equation}
 \mathbb P^1_3 \quad\rightarrow\quad\left\{
\begin{aligned}
\mathbb P^1_{3G_1}\cup \mathbb P^1_{3H_1} \qquad & \textmd{for $P_1=0$},\\
\mathbb P^1_{3G_2}\cup \mathbb P^1_{3H_2}\qquad & \textmd{for $P_2=0$},
\end{aligned}\right.                                          
\end{equation}
where $H_1$, $H_2$, $G_1$ and $G_2$ are some longish polynomials which we refrain from displaying here.
The remaining $ \pl{0},  \pl{1},  \pl{4}$ are unaffected.
 The intersection pattern of the fibred $\pl{}$'s is as in equ.~(A.31) of \cite{Krause:2011xj} and corresponds to the affine Dynkin diagram of $SU(6)$. Moreover, one readily evaluates the Cartan charges of, say, the splitting $ \pl{3}$,
 \begin{eqnarray}
\mathbb P^1_{3 G_i}: [0,1,-1,0]  = - \mu_5 + \alpha_1 + \alpha_2, \qquad \mathbb P^1_{3 H_i} = [0,0,-1,1] = \mu_5 - \alpha_1 -  \alpha_2 - \alpha_3.
\end{eqnarray}
The full identification of all weights in the ${\bf 5}$ representation is given as follows:
\begin{equation}\label{5:p1_combination}
 \begin{array}{c|c}
  \textrm{Weight}                                           &     \pl{i}-{\rm combination} \\
  \hline
  \hline
  \mu_{5}                                                                                                         &     \pl{0} +                 \pl{3 H_i} + \pl{4}        \\
  \mu_{5} - \alpha_1                                                                                      &     \pl{0} +\pl{1} +     \pl{3 H_i} + \pl{4}     \\
  \mu_{5} - \alpha_1 - \alpha_2                                                                   &     \pl{0} +\pl{1} +   \pl{2} +   \pl{3 H_i} + \pl{4}   \\
  \mu_{5} - \alpha_1-  \alpha_2 - \alpha_3                                                &     \pl{3 H_i}             \\
  \mu_{5} - \alpha_1 - \alpha_2 - \alpha_3  - \alpha_4                           &       \pl{3 H_i}   +  \pl{4}        \\
 \end{array}
\end{equation}

While the fibre structure over both ${\bf 5}$ curves is identical, the distinguishing property is again the intersection pattern with the section. Analogous considerations as for the ${
\bf 10}$ representations yield
\begin{eqnarray}
&C_{{\bf 5}^{(1)}}&   S  \cap  \mathbb P^1_{3 G_1} = 0, \qquad  S  \cap  \mathbb P^1_{3H_1} = 1, \\
&C_{{\bf 5}^{(2)}}&   S \cap  \mathbb  P^1_{3 G_1} = 1,  \qquad  S   \cap  \mathbb P^1_{3 H_2} = 0.
\end{eqnarray}

\subsubsection*{$SU(5)$-singlet curves}

There is one more type of matter curves inhabited by $U(1)_X$ charged singlets.
These extra states arise from M2-branes wrapping suitable components of the fibre over the self-intersection of the $I_1$-part of the discriminant locus. As in \cite{Grimm:2010ez}, the fibre over this self-intersection locus acquires an $SU(2)$-singularity prior to resolution. This curve of $SU(2)$ singularities is a consequence of the binomial structure of the factorised Tate mode and occurs, before the small resolution, at
$X=0 \, \cap \, Q=0 \, \cap \, Y_1 =0 \, \cap \, Y_2 =0 \subset X_5$. This describes a curve $C_{\bf 1}$ in the base space times a point $(x,t) = (x_0, t_0)$ in the fibre at which the fibre degenerates. After the small resolution (\ref{compint}), this singular point is replaced by the $\mathbb P^1$ parametrised by the homogeneous coordinates $[\lambda_1, \lambda_2]$, called ${\mathbb P}^1_{SU(2)}$ in the sequel. The original fibre, called ${\mathbb P}^1_0$, is the locus $X=0 \, \cap \, Q=0 \, \cap \, Y_1 =0 \, \cap \, Y_2 =0$ away from the point $(x_0, t_0)$, which has been blown up into ${\mathbb P}^1_{SU(2)}$. It is therefore clear that these two fibre components intersect at two points, thus forming the affine Dynkin diagram of $SU(2)$. This same structure had been discussed before in \cite{Grimm:2010ez,Braun:2011zm,Krause:2011xj,Grimm:2011fx,Morrison:2012ei}. 

Note that for the $U(1)_X$ model under consideration the curve $C_{\bf 1}$
 over which this fibre is localised is a single connected curve in the base. This follows by explicitly solving for $X=Q=Y_1=Y_2=0$ taking into account both the Stanley-Reisner-ideal and the extra restriction that $c_1=0$ and $c_0=0$ are not allowed to intersect.
 Since this curve does not lie on top of the $SU(5)$ brane its structure cannot be accounted for in any local model. This is part of the reason why a global understanding is required in the study of Abelian gauge groups. M2-branes wrapping ${\mathbb P}^1_{SU(2)}$ thus give rise to massless $SU(5)$ singlets ${\bf 1}$.

The section $S$ is given, away from the critical locus $X=0 \, \cap \, Q=0 \, \cap \, Y_1 =0 \, \cap \, Y_2 =0$ by (\ref{y1secsim}). As $X= Q=Y_1 = Y_2 =0$, however, it wraps the entire ${\mathbb P}^1_{SU(2)}$ because the homogeneous coordinates $[\lambda_1,\lambda_2]$ are now unconstrained. Therefore $S \cap {\mathbb P}^1_0 =2$ and so
\begin{eqnarray} \label{sing-S-X}
C_{\bf 1}: \quad S \cap {\mathbb P}^1_{SU(2)} = -1
\end{eqnarray}
because $S$ intersects the entire fibre class in a single point, $S \cap ({\mathbb P}^1_0+ {\mathbb P}^1_{SU(2)}) =1$.

\subsection{\texorpdfstring{$U(1)_X$}{U(1)X} charges for matter curves and \texorpdfstring{$G_4$}{G4}-flux} \label{sec:G4-41}

We now investigate the detailed relationship between the appearance of the extra section and the appearance of a $U(1)_X$.
 The relevance of $S$ lies in the fact that its dual 2-form is related to an element $\tw_X$ of $H^{1,1}(\hat Y_4)$ in terms of  which the M-theory 3-form $C_3$ can be expanded as $C_3 = A_X \wedge \tw_X + \ldots$. The 1-form $A_X$ is then the gauge potential associated with an extra $U(1)_X$ gauge symmetry, a priori in the 3-dimensional effective theory obtained by dimensional reduction of M-theory on $ \hat Y_4$. See \cite{Grimm:2010ks,Cvetic:2012xn} for recent investigations of various aspects of this effective action. 
To find the precise relation between the dual 2-form $S$\footnote{Our notation does not distinguish between a divisor $D$ and its dual 2-forms; also, the 2-form dual to divisors of the form, say, $c_1=0$ will be denoted by $c_1$.} and $\tw_X$ one requires that $\tw_X$ satisfy  the relations
\begin{eqnarray} \label{U1constraints}
&& \int_{ \hat Y_4} \tw_X \wedge D_a \wedge D_b \wedge D_c = 0, \qquad \int_{\hat Y_4} \tw_X \wedge Z \wedge D_a \wedge D_b = 0 ,\\
&& \int_{ \hat Y_4} \tw_X \wedge E_i \wedge D_a \wedge D_b =0,  \qquad i={1,\ldots,4}
\end{eqnarray}
with $D_a, D_b, D_c$ the pullback of arbitrary base divisors. The first two constraints ensure that under F/M-theory duality $A_X$ actually lifts to a 1-form in four dimensions. The last constraint normalises the $U(1)_X$ generator to be orthogonal to the Cartan generators of the non-Abelian gauge group, in our case $SU(5)$. The first two constraints were worked out for the $U(1)$ restricted Tate model in \cite{Grimm:2010ez} (see also \cite{Braun:2011zm}) and for the $SU(5) \times U(1)_X$ restricted Tate model with a single ${\bf 10}$-matter curve in \cite{Krause:2011xj} (see also \cite{Grimm:2011fx}). In the mathematics literature the map from $S$ to $\tw_X$ is known as the Shioda map \cite{Shioda1}, as reviewed recently e.g. in \cite{Morrison:2012ei,Cvetic:2012xn}. 

As for the first constraint, observe that 
\begin{eqnarray}
 \int_{\hat Y_4} S \wedge D_a \wedge D_b \wedge D_c = \int_B D_a \wedge D_b \wedge D_c
\end{eqnarray}
because $S$ is a section. Thus we subtract $Z$ because  $\int_{\hat Y_4} (S-Z) \wedge D_a \wedge D_b \wedge D_c =0$.
Next we compute 
\begin{eqnarray}
\int_{\hat Y_4} (S-Z) \wedge Z \wedge D_a \wedge D_b = \int_{\hat Y_4} S \wedge Z \wedge D_a \wedge D_b + \int_B \bar{\cal K} \wedge D_a \wedge D_b,
\end{eqnarray}
where we used that $Z Z =  - Z \bar{\cal K}$ in terms of the anti-canonical class $\bar{\cal K}$ of $B$.
The intersection $\int_{\hat Y_4} S \wedge Z \wedge D_a \wedge D_b$ is evaluated in the complete intersection $\hat X_6$ as the number of generic intersections of 
\begin{eqnarray}
x=t^2  \quad \cap \quad  c_1 t = 0  \quad \cap \quad        \lambda_1 = 0 \quad \cap \quad z=0     \quad \cap \quad     D_a = 0    \quad \cap \quad     D_b    =0
\end{eqnarray}
in $\hat X_6$. The first two constraints are $Y_1=0$ and $X=0$ evaluated for $z=0$. Since $x=t=z=0$ is excluded by the Stanley-Reisner ideal, the intersection is
\begin{eqnarray}
\int_{\hat Y_4} S \wedge Z \wedge D_a \wedge D_b = \int_{\hat Y_4} c_1 \wedge Z \wedge D_a \wedge D_b = \int_B c_1 \wedge D_a \wedge D_b.
\end{eqnarray}
Thus the first two constraints are satisfied by\footnote{We hope the reader is not confused by the fact that $c_1$ denotes the Tate polynomial in $a_1= d_4 c_1$ and not, as oftentimes in the literature, $c_1(B)$. We will always express $c_1(B)$ in terms of $\bar{\cal K}$.} $S - Z - \bar {\cal K} - c_1$.

Finally we implement  $\int_{\hat Y_4} \tw_X \wedge E_i \wedge D_a \wedge D_b =0$.
Since $z e_i$ is in the Stanley-Reisner ideal the only constraint arises from the intersection with $S$, which is given by
\begin{eqnarray}
\int_{\hat Y_4} S \wedge E_j \wedge D_a \wedge D_b = \delta_{j3} \int_B W \wedge D_a \wedge D_b.
\end{eqnarray}
To eliminate this intersection with $E_3$ without spoiling the first two constraints we add a linear combination $\sum_{i=1}^4 t_i E_i$ such that $\sum_i t_i \int_{ \hat Y_4} E_i \wedge E_j \wedge D_a \wedge D_b = \sum_i  t_i C_{ij} \int_B W \wedge D_a \wedge D_b = - \delta_{j3}$ with $C_{ij}$ the $SU(5)$ Cartan matrix, cf. eq. (\ref{Cij}).  
In total the correct $U(1)_X$ generator is 
\begin{eqnarray}
\tw_X = 5(S - Z - \bar{\cal K} - c_1) + \sum_i t_i E_i, \qquad \quad t_i = (2,4,6,3).
\end{eqnarray}
Here we have picked the overall normalisation of $\tw_X$ such that no factional charges will appear.
Note that for $c_1\equiv 1$ this reduces to the expression found in  \cite{Krause:2011xj} for the $SU(5) \times U(1)_X$ restricted Tate model with one ${\bf 10}$-curve.

We are now in a position to compute the $U(1)_X$ charges of the ${\bf 10}$ representation localised on the the matter curves.
These are given by
\begin{eqnarray}
q_X = \int_{\sum \pl{ij}} \tw_X
\end{eqnarray}
with $\sum \pl{ij}$ denoting the linear combination of $\mathbb P^1$s in the fibre of the respective matter curves corresponding to one component of the weight vector of the representation. Of course  the value of the integral is the same for all weights. 
This expression is easiest computed for the weight $\mu_{10}- \alpha_2 -\alpha_3$ corresponding to $\pl{03}$. 
The integral $\int_{\mathbb P^1_{03}} E_i$ gives just the Cartan charges $[1,0,-1,1]$ of this weight  and thus $\int_{\mathbb P^1_{03}}\sum_i t_i E_i = -1$. Furthermore $z e_3$ is in the Stanley-Reisner ideal so that $\int_{\mathbb P^1_{03}}Z =0$, as is  $\int_{\mathbb P^1_{03}}- \bar{\cal K} - c_1$\footnote{Consider first $\int_{\pl{03}} \bar {\cal K}$: For the fibre over $d_4=0$  this intersection is $e_0 = 0   \cap  e_3=0   \cap d_4 = 0 \cap D_a= 0  \cap \bar{\cal K}=0 \subset \hat X_5$ for an arbitrary divisor in the base $D_a$ that intersects the matter curve once. This vanishes because on the 4-fold, $e_0=0$ is constrained to lie over the $SU(5)$ divisor $w=0$ in the base, and the generic intersection of this with the three more base divisors $D_a=0, d_4=0$ and $\bar {\cal K}$ vanishes. The same holds for $\int_{\pl{03}} \bar {\cal K}$ over $c_1=0$. By a similar argument $\int_{\pl{03}} c_1 =0$ over $d_4=0$. Over $c_1=0$ on the other hand,  $\int_{\pl{03}} c_1 $  boils down to $\int_{\hat Y_4} E_0 \wedge E_3\wedge c_1 \wedge D_a =  C_{03} \int_B W \wedge D_a \wedge c_1 = 0$ with $C_{03}=0$ the corresponding entry from the extended Cartan matrix  of $SU(5)$.}.
Now it becomes crucial that the intersection pattern of $\mathbb P^1_{03}$ with $S$ differs for the fibre over the two ${\bf 10}$ curves as given in (\ref{ScapP}). Adding up these contributions yields
\begin{eqnarray}
C_{{\bf 10}^{(1)}}: \quad   q_{{\bf 10}^{1}} = -1, \quad \qquad  C_{{\bf 10}^{(2)}}: \quad  q_{{\bf 10}^{(2)}} = 4.
\end{eqnarray}

A similar computation for the {\bf 5}-curves leads to the charges
\begin{eqnarray}
C_{{\bf 5}^{(1)}}: \quad   q_{{\bf 5}^{(1)}} = 2, \quad \qquad  C_{{\bf 5}^{(2)}}: \quad  q_{{\bf 5}^{(2)}} = -3.
\end{eqnarray}

Finally, the singlets from M2-branes wrapping $\mathbb P^1_{SU(2)}$ have charge $- 5$ as a consequence of (\ref{sing-S-X}) and the fact that $Z$ and $E_i$ have zero intersection with $\mathbb P^1_{SU(2)}$,
\begin{eqnarray}
C_{\bf 1}: \quad q_{\bf 1} = -5.
\end{eqnarray}
The conjugate singlets are due to M2-branes wrapping $\mathbb P^1_0$. This is the component of the singular $SU(2)$ fibre intersected by the universal section, $Z \cap {\mathbb P}^1_0 =1$ so that $\int_{{\mathbb P}^1_0} \tw_X = 5$ because $S \cap \mathbb P^1_0 =2$.

To conclude this section we stress that as in \cite{Grimm:2010ez,Braun:2011zm,Krause:2011xj,Grimm:2011fx} the extra $U(1)_X$ gauge group opens up the possibility of switching on associated non-trivial gauge flux. By F/M-theory duality, such gauge flux is described in terms of the M-theory 4-form field strength
\begin{eqnarray}
G_4^X = F \wedge \tw_X, \qquad F \in H^{1,1}(B).
\end{eqnarray}
In particular this induces a chiral spectrum of charged matter states $R_i$ with chiral index given by
\begin{eqnarray}
\int_{{\cal C}_{R_i}} G^X_4 = q_i \int_{C_{R_i}} F,
\end{eqnarray}
where the states $R_i$ are localised on the matter surface ${\cal C}_{R_i}$, which is $\mathbb P^1$-fibred over the curve $C_{R_i}$ in the base $B$.
With the help of our results for the charges $q_i$ the computation of the chiral index thus reduces to evaluating the integral of the flux $F$ over the matter curve in the base.

\subsection{Yukawa points}

We now come to the points of Yukawa interactions at the intersection of the various matter curves.

An interesting feature compared to the previously analysed $U(1)$-restricted Tate model with only a single ${\bf 10}$-curve is that the intersection of the two ${\bf 10}$ distinct matter curves gives rise to the Yukawa coupling ${\bf 10}^{(1)}_{-1} \, {\bf \overline{10}}^{(2)}_{-4} {\bf 1}_5$.
This field theoretic expectation is confirmed by an explicit analysis of the fibre structure over the intersection of $C_{{\bf 10}^{(1)}}$ and $C_{{\bf 10}^{(2)}}$  along the $SU(5)$ divisor in the base, corresponding to $c_1=d_4=w=0$. To this end we can start from the hypersurface equations (\ref{P1-hyperA}) and set $c_1=0$.
In particular, $\mathbb P^1_{3C_1}$ splits as 
\begin{eqnarray}
P^1_{3C_1} \rightarrow \mathbb P^1_{03} \cup \pl{43} \cup \pl{3\tilde C}
\end{eqnarray}
with the polynomial $\tilde C = (d_0 e_0^2 e_1 e_4 + d_2 x) + d_3 y$.
This leaves us with six $\mathbb P^1$s (including multiplicities),
\begin{eqnarray} \label{P1-10-10-1}
1 \times \pl{3 \tilde C}, \quad 1\times \pl{2\tilde B}, \quad 1 \times \pl{0A}, \quad 2 \times \pl{34}, \quad 2 \times \pl{13}, \quad 3 \times \pl{03}
\end{eqnarray}
with $\tilde B = e_3- c_0 d_3 e_1 z^3$.
The intersection structure of these $\mathbb P^1$s follows again by counting common solutions of the involved hypersurfaces with the help of the Stanley-Reisner ideal (\ref{srideal_optional}). For example, the intersection $\pl{34} \cap \pl{0A}  $ is characterised by the \emph{six} equations
\begin{eqnarray}
e_3 =0 \quad \cap \quad e_4=0 \quad \cap \quad e_0 \quad \cap \quad e_1=e_4 \quad \cap \quad d_4=0 \quad \cap \quad c_1=0
\end{eqnarray}
inside the ambient 5-fold $\hat X_5$, which generically has no solution. In other cases some of the constraints coincide and common solutions are possible.
This way one can establish that the six $\mathbb P^1$s intersect with one another like the (non-extended) Dynkin diagram of $E_6$.

So far we have not taken into account the $SU(2)$ singularity and its resolution.
Recall from (\ref{ScapP}) that over generic points on the ${\bf 10}$ matter curves $C_{{\bf 10}^{(1)}}$ and $C_{{\bf 10}^{(2)}}$, the resolved section $S$ intersects $\pl{3C_1}$ and $\pl{03}$, respectively. The transition between both intersections occurs as the two matter curves intersect, where $\pl{3C_1}$ splits off an extra copy of $\pl{03}$.
Indeed, at the intersection of the locus $c_1=0 \cap d_4=0 \cap e_0=0 \cap e_3 =0$  with $x=t^2$, the polynomials $Y_1, Y_2, X, Q$ all vanish, indicating that this point lies on the $SU(2)$ singular locus which is resolved by the small resolution (\ref{compint}). The singular point is replaced by the $\mathbb P^1_{SU(2)}$ described by the homogeneous coordinates $[\lambda_1, \lambda_2]$. Therefore the central $\mathbb P^1_{03}$ of the $E_6$ Dynkin diagram discussed above is intersected by this extra $\mathbb P^1_{SU(2)}$. 
The topology of the fibre including this incoming $\mathbb P^1_{SU(2)}$ is depicted in figure \ref{fig:10101-coupling-4-1-case}. Note that the specific intersection pattern may depend on the concrete choice of the Stanley-Reisner ideal, i.e.\ on the particular  triangulation one is working with.
\begin{figure} 
 \centering
 \def\svgwidth{0.65\linewidth}
 \executeiffilenewer{10101-coupling-4-1-case.svg}{10101-coupling-4-1-case.pdf}%
 {inkscape -z -D --file=10101-coupling-4-1-case.svg %
  --export-pdf=10101-coupling-4-1-case.pdf --export-latex}%
   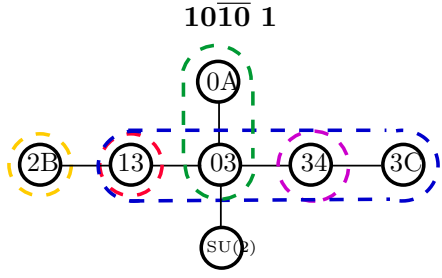%

 \caption{Schematic drawing of the  ${\bf 10}^{(1)} \, {\bf \overline{10}}^{(2)}{\bf 1} $ Yukawa coupling in the case of the 4-1 split.}
\label{fig:10101-coupling-4-1-case}
\end{figure}

For the intersection of the singlet with the two {\bf 5}-curves, i.e.~the $\overline{\mathbf 5}^{(1)}_{-2} \mathbf 5^{(2)}_{-3} \mathbf 1_{5}$ Yukawa coupling, we obtain the same $SU(7)$ pattern as in~\cite{Krause:2011xj}. Here, as in~\cite{Krause:2011xj}, the $\pl{}$ corresponding to the resolution of the $SU(2)$-singularity `appears' between $\pl{3G_1}$/$\pl{3G_1}$ and $\pl{3H_1}$/$\pl{3H_2}$. The same methods also allow for an analogous analysis of the remaining familiar Yukawa coupling points between $SU(5)$ charged matter states.

\section{Fibre Structure and charges in the \texorpdfstring{$U(1)_{PQ}$}{U(1)PQ} model}
\label{sec:fibu1pq}

We now address the $3-2$ factorised Tate mode of section \ref{sec:32split}  in more detail. The additional $U(1)$ is referred to as of  Peccei-Quinn type and denoted by $U(1)_{PQ}$ in the local model building literature. This corresponds to the fact that for a local $3-2$ split it is possible to assign the Higgs up and down multiplets to different matter curves \cite{Marsano:2009wr} and so they have different charges under $U(1)_{PQ}$.

\subsubsection*{10-curves}
Let us begin with the two {\bf 10}-matter curves
\begin{eqnarray}
C_{{\bf 10}^{(1)}}:  \quad  d_3 = 0 \quad \cap \quad w=0, \qquad \quad C_{{\bf 10}^{(2)}}:  \quad c_2 = 0 \quad \cap  \quad w=0.
\end{eqnarray}
An analysis of the $\mathbb P^1$ split completely analogous to that in section \ref{sec:fibstruc} yields a fibre structure over both curves identical to the one given in (\ref{10_1:p1_combination}), even though, of course, the explicit form of the polynomials $B_1, C_1, D_1$ and $B_2, C_2 ,D_2$ differs. 
What is very interesting, on the other hand, are the changes in the intersection structure of $\pl{03}, \pl{13}$ and $\pl{3C_i}$ with the section $S$. This is of course crucial to determine the correct $U(1)$ charges.

The extra section $S$ is still given, over generic loci, by $\lambda_1= 0  \cap  X =0  \cap  Y_1 =0 \subset \hat X_6$, but with 
\begin{eqnarray}
Y_1 = c_2 t^2 + \alpha \delta e_0 t z + \alpha \beta e_0^2 z^2, \qquad X = t^2 e_4 - x.
\end{eqnarray}
The big difference to the $4-1$ model is that $Y_1$ is a polynomial of degree 2, not of degree 1, in $t$. This changes the intersection pattern as follows:
First note that the intersection of $S$ with the resolution $\mathbb P^1_3$ over a generic point on the $SU(5)$ divisor $w=0$ is (setting $z =1$ and $e_4=1$)
\begin{eqnarray}
e_3 = 0 \quad \cap \quad \lambda_1=0 \quad \cap \qquad x = t^2 \quad \cap \quad Y_1=0 \quad \cap \quad D_a=0 \quad \cap \quad D_b =0
\end{eqnarray}
inside $\hat X_6$ with $D_a$ and $D_b$ intersecting $w=0$ once in the base.
Since $Y_1$ is of degree 2, the intersection number is
\begin{eqnarray} \label{S3b}
\mathbb P^1_3  \cap  S = 2.
\end{eqnarray}

Let us now compute the intersections with the descendants of ${\mathbb P^1_3}$ over the ${\bf 10}$ curves, beginning with $C_{{\bf 10}^{(2)}}$.
Concerning $\pl{03}$ we observe that $Y_1|_{c_2=e_0=e_3 =0} =0$ so that, as is the case for  $C_{{\bf 10}^{(2)}}$ in section \ref{sec:fibstruc}, ${\pl{03}} \cap S = 1$.  Unlike before, however, also ${\pl{3C_2}} \cap S = 1$ because
$Y_1|_{c_2=e_3=0} =  \alpha \delta e_0 t + \alpha \beta e_0^2$ (after setting $z=1$) and if we solve this degree 1 polynomial for $t$ and plug the solution into $C_2$, the latter vanishes automatically along $\pl{3C_2}$.  Note that indeed $\pl{03} \cap S +  \pl{3C_2} \cap S =  \pl{3} \cap S$.

Over $C_{{\bf 10}^{(1)}}$, corresponding to $d_3=0$,  $\pl{03} \cap S =0$ as no simplifications in $Y_1$ occur. To compute the intersection $\pl{3C_1}$ we note that solving $Y_1|_{e_3=0}=0$ for $t$ gives us two solutions because $Y_1$ is degree 2. For each of these, $C_1$ vanishes automatically once we impose all other constraints of the defining equation of $\pl{3C_1}$. Thus ${\pl{3C_1}} \cap S =2$. To summarise,
\begin{eqnarray}\label{ScapP32}
&C_{{\bf 10}^{(1)}}:&   S \cap  \mathbb P^1_{03} = 0,  \qquad  S   \cap  \mathbb P^1_{13} = 0, \qquad    S  \cap  \mathbb P^1_{3C_1} = 2, \nn \\
&C_{{\bf 10}^{(2)}}:&   S  \cap  \mathbb P^1_{03} = 1, \qquad  S  \cap  \mathbb P^1_{13} = 0, \qquad S  \cap  \mathbb P^1_{3C_2} = 1.
\end{eqnarray}

\subsubsection*{5-curves}

There are now three  ${\bf 5}$ curves located at the intersection of the $SU(5)$ divisor with the zero locus of the three polynomials $P_1, P_2, P_3$ into which $P=a_1^2 a_{6,5} - a_1 a_{3,2} a_{4,3} + a_{2,1} a_{3,2}^2$ factorises,
\begin{eqnarray}
&& P_1 = \delta, \qquad P_2 = \beta d_3 + d_2 \delta, \\
&&  P_3 = \alpha^2 c_2 d_2^2 + \alpha^3 \beta d_3^2 + \alpha^3 d_2 d_3 \delta - 2 \alpha c_2^2 d_2 \gamma - 
  \alpha^2 c_2 d_3 \delta \gamma + c_2^3 \gamma^2.
\end{eqnarray}
In the fibre over each of these three matter curves 
\begin{eqnarray}
C_{{\bf 5}^{(i)}}: \quad P_i = 0 \quad \cap \quad w=0 \subset B, \quad i=1,2,3
\end{eqnarray}
 one observes that $\mathbb P^1_3$ splits according to
\begin{eqnarray}
\mathbb P^1_3 \rightarrow \mathbb P^1_{3 G_i} \cup \mathbb P^1_{3 H_i}.
\end{eqnarray}
The explicit form of $G_i$ and $H_i$ is rather lengthy, in particular for $i=3$. However, in all three cases one can easily evaluate the Cartan charges of $\mathbb P^1_{3 G_i}$ and of $\mathbb P^1_{3 H_i}$ as
\begin{eqnarray}
\mathbb P^1_{3 G_i}: [0,1,-1,0]  = - \mu_5 + \alpha_1 + \alpha_2, \qquad \mathbb P^1_{3 H_i} = [0,0,-1,1] = \mu_5 - \alpha_1 -  \alpha_2 - \alpha_3.
\end{eqnarray}
In fact, in all three cases the complete weight assignments of the ${\bf 5}$ representation are exactly as in (\ref{5:p1_combination}).

What distinguishes the three ${\bf 5}$ curves is the intersection pattern of the respective fibres with the section $S$ and thus the $U(1)_{PQ}$ charges. 
An explicit analysis of the defining polynomials of all the $\mathbb P^1$s reveals
\begin{eqnarray}
&& \pl{3G_1} \cap S = 2, \qquad \pl{3H_1} \cap S = 0, \\
&& \pl{3G_2} \cap S = 0, \qquad \pl{3H_2} \cap S = 2, \\
&& \pl{3G_3} \cap S = 1, \qquad \pl{3H_3} \cap S = 1. 
\end{eqnarray}
The logic behind these computations is identical to the ${\bf 10}$-curves:  Concerning $\pl{3G_1}$, one can solve the quadratic polynomial $Y_1|_{e_3=0}$ for $t$ and confirm that for both solutions in $t$ the polynomial $G_1$ vanishes identically if we take into account all further polynomials entering the section $S$ and $\mathbb P^1_{3G_1}$. By contrast for $\pl{3H_1}$ no such simplifications occur and thus the intersection number vanishes. 
On the other hand, for $\pl{3G_3}$ it is simpler to solve $G_3$ for $x$ and combine this with $x=t^2$ into an equation for $t$, again with two solutions. Crucially, only one of these solves $Y_1=0$, leading to $\pl{3G_3} \cap S = 1 = \pl{3H_3} \cap S$.

\subsubsection*{Singlet curves}

The singlet curves in the $U(1)_{PQ}$ model are particularly interesting and exhibit additional structure. In fact we encounter 3 types of singlets and to see how these types are classified it is worth discussing in more detail the loci on which singlets are expected to localise. Consider the section 
\begin{equation}
X = Y_1 = 0 \;.
\end{equation} 
Because $Y_1$ is a quadratic polynomial this defines two points on the torus which are the two roots of $Y_1=0$,
\begin{equation}
t = \frac{e_0 z}{2c_2} \left( -\alpha \delta \pm \sqrt{- 4 \alpha \beta c_2 + \alpha^2 \delta^2} \right) \;.
\end{equation} 

The first type of singlets are the usual ones as in the $U(1)_X$ case. These correspond to loci where a single root of $Y_1$ coincides with a single root of $Y_2$. 
Recall that prior to resolution, the Tate model is singular along the curve $X= Q= Y_1= Y_2 = 0$, and these singlets localise on the locus  $C_{{\bf 1}^{(1)}}$ corresponding to the \emph{generic} solution of these four polynomials. As will become clear momentarily, this is the locus away from $(x,t) = (0,0)$,
\begin{eqnarray}
& C_{{\bf 1}^{(1)}}:  & X= 0 \quad \cap \quad Q=0 \quad \cap \quad Y_1 = 0 \quad \cap \quad Y_2 = 0\;,  \label{su2curve}   \quad {\rm} \quad (x,t) \neq (0,0).    \label{PQ-C1}
\end{eqnarray}
 After resolution the section $S$ wraps the resolution $\mathbb P^1_{SU(2)}$ in the fibre over  $C_{{\bf 1}^{(1)}}$ and by the same arguments as in the $U(1)_X$ model the intersection number is
 \begin{eqnarray}
 C_{{\bf 1}^{(1)}}: \qquad   {\mathbb P}^1_{SU(2)} \cap S = -1.%
 \end{eqnarray}
This leads to singlets with charge $\pm 5$ which localise on {\it generic} solutions to (\ref{su2curve}).

The second type of singlets are again charged ones that localise on special sub-loci of the curve (\ref{su2curve}) where additionally two roots in the $Y_1$ and two roots in the $Y_2$ factors degenerate so that in total four roots degenerate. The loci where this occurs are
\begin{eqnarray}
C_{{\bf 1}^{(2)}}:  & x= 0 \quad \cap \quad t=0 \quad \cap \quad \beta = 0 \quad \cap \quad \delta = 0\;, \label{spec2} \\
C_{{\bf 1}^{(3)}}:  & x= 0 \quad \cap \quad t=0 \quad \cap \quad \gamma = 0 \quad \cap \quad \alpha = 0\;. \label{spec3}
\end{eqnarray}
As discussed in section \ref{sec:factate}, because these loci are at $x=0$ we should check that the expected $SU(2)$ singularity is present also in the coordinates $\{x,y\}$, which indeed can be confirmed. Already from the previous discussion we can guess the charges of these states: Because charge $\pm 5$ singlets are localised where one root from $Y_1$ coincides with one root from $Y_2$, here we expect states of the double charge, i.e.\ with $q= \pm 10$. Due to the different form of the binomial singularity on this locus, and potential subtleties on the locus $x=0$, we will not perform the resolution on these special loci in the current approach. Instead we will explicitly show how to recover the doubly charged singlets associated to $C_{{\bf 1}^{(2)}}$ in a different formalism in section \ref{sec:relmorpa}.\footnote{\label{footnote1}The arguments just given suggest similar results for $C_{{\bf 1}^{(3)}}$, but our analysis of section \ref{sec:relmorpa} is valid only if $\gamma = 0 \cap \alpha=0$ is empty 
so more work is required in cases where this constraint is not met. In practice we can bypass this problem by restricting ourselves to base spaces $B$ such that $\alpha=0$ and $\gamma=0$ do not intersect, and this is the approach we are going to take from now on.}

The curve (\ref{su2curve}) also has other special solutions analogous to (\ref{spec2})-(\ref{spec3}) but where only three roots, rather than four, degenerate. For example the loci $x=t=\alpha=\beta=0$ and $x=t=\gamma=\beta=0$. However it can be checked that the manifold is not singular on these loci by directly analysing the Tate polynomial (\ref{tateform}) in the coordinates $\{x,y\}$. More precisely the singular loci are on the locus $a_{6,5}=a_{4,3}=a_{3,2}=0$ where
\begin{eqnarray}
a_{6,5}&=&\alpha \beta^2 \gamma \;, \nn \\
a_{4,3}&=& \alpha \beta d_2 + \beta c_2 \gamma - \alpha \delta^2 \gamma \;, \nn \\
a_{3,2}&=& \alpha \beta d_3 + \alpha d_2 \delta - c_2 \delta \gamma \;.
\end{eqnarray}  

The third type of singlets are associated to loci where a degeneration of roots inside the same factor $Y_1$ occurs. We expect completely neutral singlets to localise there but since the manifold is non-singular on this locus this is harder to show explicitly. Note that the locus where the roots degenerate $4\beta c_2 = \alpha \delta^2$ can be written as 
\begin{equation}
4 c_0 c_2 = c_1^2 \;. \label{singcur}
\end{equation} 
The projection of this curve to the GUT brane $w=0$ was indeed identified in \cite{Marsano:2009wr} as the expected projection of the neutral singlets from the group theory.

\subsubsection*{$U(1)_{PQ}$ generator and charges}

The generator $\tw_{PQ}$ of the $U(1)_{PQ}$ symmetry is determined by the same logic as in section \ref{sec:G4-41}. 
What differs is first that $\int_{\hat Y_4} S \wedge Z \wedge D_a \wedge D_b  = \int_B c_2 \wedge D_a \wedge D_b$ and second that the intersection number with $E_3$ is now 2, not 1, see (\ref{S3b}).  This fixes $\tw_{PQ}$ to take the form
\begin{eqnarray}
\tw_{PQ} = 5(S - Z - \bar{K} - c_2) + 2 \sum_i t_i E_i, \qquad \vec{t}=(2,4,6,3),
\end{eqnarray}
again for a convenient choice of overall normalisation. 

Consequently we arrive at the following $U(1)_{PQ}$ charges:
\begin{eqnarray}
&& C_{{\bf 10}^{(1)}}: \quad   q_{{\bf 10}^{(1)}} = -2, \quad \qquad  C_{{\bf 10}^{(2)}}: \quad   q_{{\bf 10}^{(2)}} = 3, \\
&& C_{{\bf 5}^{(1)}}: q_{{\bf 5}^{(1)}} = - 6, \qquad    C_{{\bf 5}^{(2)}}: q_{{\bf 5}^{(2)}} =   4, \qquad C_{{\bf 5}^{(3)}}: q_{{\bf 5}^{(3)}} =   -1, \\
&& C_{{\bf 1}^{(1)}}: q_{{\bf 1}^{(1)}} = -5.
\end{eqnarray}
The remaining singlets over (\ref{spec2}) will be discussed in section \ref{sec:relmorpa} and have charges (see footnote {\ref{footnote1}}) 
\begin{equation}
C_{{\bf 1}^{(2)}}: q_{{\bf 1}^{(2)}} = 10.
\end{equation} 

\subsubsection*{Yukawa points}

The Yukawa coupling ${\bf 10}^{(1)}_{-2} {\bf  \overline{10}}^{(2)}_{-3} {\bf 1}^{(1)}_5$ (with the subscripts denoting $U(1)_{PQ}$ charge for clarity) is located at the triple intersection of $C_{{\bf 10}^{(1)}}$, $C_{{\bf 10}^{(2)}}$ and the generic locus (\ref{PQ-C1}).  The same splitting as in (\ref{P1-10-10-1}) occurs, the only difference being that the section $S$ intersects both $\pl{03}$ and $\pl{3 \tilde C}$ once - at least in the current triangulation used. This modifies the intersection pattern of the fibre as given in figure \ref{fig:10101-coupling-3-2-case}. 
\begin{figure} 
 \centering
 \def\svgwidth{0.65\linewidth}
 \executeiffilenewer{10101-coupling-3-2-case.svg}{10101-coupling-3-2-case.pdf}%
 {inkscape -z -D --file=10101-coupling-3-2-case.svg %
  --export-pdf=10101-coupling-3-2-case.pdf --export-latex}%
   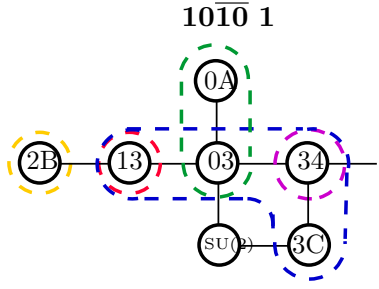%

 \caption{Schematic drawing of the $\mathbf{10}^{(1)}$ $\mathbf{\overline{10}}^{(2)}$ $\mathbf{1}$ Yukawa coupling in the case of the 3-2 split.}
\label{fig:10101-coupling-3-2-case}
\end{figure}

The locus of the doubly charged singlets $C_{{\bf 1}^{(2)}}$ intersects two of the {\bf 5}-matter curves and the charges precisely agree with an associated cubic interaction
\begin{equation}
{\bf 5}^{(1)}_{-6} \, \bar{\bf 5}^{(2)}_{-4} \, {\bf 1}^{(2)}_{10} \;.
\end{equation} 
 It can be checked, in the framework of section \ref{sec:relmorpa}, that indeed the fibre exhibits the structure of an $SU(7)$ enhancement at that point.

\subsubsection*{Implications for group theoretic embedding into $E_8$}

The appearance of the extra charged singlets ${\bf 1}^{(2)}_{\pm 10}$ is quite surprising from a group theoretic perspective: The common lore in the literature is that global Tate models are based on a \emph{single} $E_8$ gauge group which is broken to the gauge group $G \subset E_8$ along the divisor $w=0$. As reviewed in appendix \ref{App-SCC} this breaking can be understood locally in terms of a Higgs bundle  \cite{Donagi:2009ra} with structure group $SU(5)_\perp$, which in the present case would factorise into $SU(3)_{\perp} \times SU(2)_{\perp} \times U(1)_{PQ}$.
However, the decomposition of ${\bf 248}$ of a single $E_8$ into irreducible representations of $SU(5) \times SU(3)_{\perp} \times SU(2)_{\perp} \times U(1)_{PQ}$ only gives rise to a single type of charged singlets ${\bf 1}_{\pm 5}$. The appearance of two types of charged singlets  ${\bf 1}^{(1)}_{\pm 5}$ and ${\bf 1}^{(2)}_{\pm 10}$ in the factorised $U(1)_{PQ}$ Tate model implies that no embedding into a \emph{single} underlying $E_8$ is possible. This conclusion is also supported by the structure of the Yukawa couplings:
A further specialisation of the complex structure moduli of the 4-fold can lead to points of $E_8$ enhancements on the divisor $w=0$ where the Yukawa couplings
${\bf 10}^{(2)}  {\bf 10}^{(2)}   {\bf 5}^{(1)}$,  ${\bf 10}^{(2)}  {\bf \overline 5}^{(2)}   {\bf 5}^{(2)}$ and ${\bf 5}^{(1)} \bar{\bf 5}^{(3)}  {\bf 1}^{(1)} $ come together.
This corresponds to the embedding of all these representations into a single $E_8$. By contrast, the Yukawa coupling ${\bf 5}^{(1)}  \bar{\bf 5}^{(2)}  {\bf 1}^{(2)}$ can never coincide with this point of $E_8$ as this would require that $c_2= \beta = \delta = 0$ on a single point on $w=0$, but the intersection of $c_2 \cdot \beta  \cdot  \delta$  must be forbidden as it would lead to a non-Kodaira singularity, see eq. (\ref{3-2-Kodaira}).

Put differently, if one were to construct a heterotic dual of a $3-2$ factorised model with $w=0$ the base of a K3-fibration, the heterotic dual would have to be singular in such a way as to incorporate the extra charged singlets ${\bf 1}^{(2)}_{\pm 10}$  in a non-perturbative fashion as  these cannot arise from the same perturbative heterotic $E_8$ factor as the remaining states.

\section{Factorised \texorpdfstring{$SU(5) \times U(1)$}{SU(5) x U(1)} models as \texorpdfstring{$\mathbb P_{[1,1,2]}$}{P[1,1,2]}-fibrations}
\label{sec:relmorpa}

In this section we provide a rather different, but equivalent description of the factorised Tate models with one extra $U(1)$-symmetry. 
Motivated by a study of  the landscape of 6-dimensional F-theory compactifications, ref.~\cite{Morrison:2012ei} recently provided the general form of a Weierstra\ss{} equation that describes an elliptic fibration with two independent sections over the 2-complex dimensional base space $\mathbb P^2$. The conclusion of~\cite{Morrison:2012ei} is that such a Weierstra\ss{} model can be written as
\begin{equation}\label{eq:rest-weierstrass}
 \begin{split}
 Y^2=X^3&+\left(\C_1 \,\C_2-\B^2\, \C_0-\tfrac13\, \C_2^2\right)\,X\,Z^4+\\
&+\left(\C_0\,\C_3^2-\tfrac13 \,\C_1 \,\C_2\, \C_3+\tfrac{2}{27}\, \C_2^3-\tfrac23 \,\B^2\, \C_0\, \C_2+\tfrac14 \B^2\, \C_1^2\right)\,Z^6\,. 
 \end{split}
\end{equation}
Here the fibre coordinates $X,Y,Z$ are homogeneous coordinates on $\mathbb P_{[2,3,1]}$ and $\B$ and $\C_i$ denote some generic sections of some line bundles over the base, which in the case of~\cite{Morrison:2012ei} was taken to be $\mathbb P^2$. Indeed, for a Weierstra\ss{} model of the form~\eqref{eq:rest-weierstrass} one finds that
\begin{equation}\label{eq:section-MP}
[X,Y,Z]=[\C_3^2-\tfrac23 \, \B^2 \, \C_2,\,-\C_3^3+\B^2 \, \C_2 \, \C_3-\tfrac12 \, \B^4 \, \C_1,\,\B]
\end{equation}
solves the Weierstra\ss{} equation and therefore represents an additional section besides the universal zero section $Z=0$.

For the same reasons as in the factorised Tate models the restriction (\ref{eq:rest-weierstrass}) of the Weierstra\ss{} model responsible for this extra section renders the model singular in codimension 2. In \cite{Morrison:2012ei} these singularities are resolved by translating the $\mathbb P_{[2,3,1]}$-fibration into a $\mathbb P_{[1,1,2]}$ fibration with homogeneous coordinates $\textmd w$, $\textmd v$, $\textmd u$ and then blowing up the point $\textmd w=\textmd u=0$ in the fibre. This introduces a blow-up divisor with coordinate $s$. The resolved space then takes the form of  a $\textmd{Bl}_{[0,1,0]}\mathbb P_{[1,1,2]}$-fibration
\begin{equation}\label{eq:211-hse-ng}
  \B\,\textmd v^2\, \textmd w +  s\, \textmd w^2  =\C_3\,\textmd v^3\,\textmd  u +  \C_2\, s \, \textmd v^2 \,\textmd  u^2 +  \C_1 \, s^2 \,\textmd v\,\textmd  u^3 +   \C_0\, s^3\,\textmd  u^4\,,
\end{equation}
with $\textmd w$, $\textmd v$, $\textmd u$ and $s$ the homogeneous coordinates of $\textmd{Bl}_{[0,1,0]}\mathbb P_{[1,1,2]}$.
Indeed this blow-up procedure for the fibre over the $SU(2)$ singular curve and the resulting transition to a $\textmd{Bl}_{[0,1,0]}\mathbb P_{[1,1,2]}$ fibration had also been applied in the $U(1)$ restricted Tate model \cite{Morrison:2012ei}, which is a special case of the model (\ref{eq:211-hse-ng}). See furthermore \cite{Aldazabal:1996du,Aluffi:2009tm} for the relevance of different fibration-types in F-theory compactifications.

Since the results of~\cite{Morrison:2012ei} apply to any fibration with two independent sections, it must be possible to bring the factorised Tate models with just one $U(1)$ symmetry into the form (\ref{eq:rest-weierstrass}).
More precisely, the $SU(5) \times U(1)_X$ and the $SU(5) \times U(1)_{PQ}$ models of sections \ref{sec:fibstruc-gen} and \ref{sec:fibu1pq} should arise as  further specialisations of  (\ref{eq:rest-weierstrass}) such as to account for the non-Abelian gauge symmetry along $w=0$.

In the case of the $4-1$ factorisation, it is indeed straightforward to provide the identification with~\eqref{eq:rest-weierstrass}. For $\C_1$, $\C_2$, $\C_3$ and $\B$ we only have to consider the definition of the section that we gave in section~\ref{sec:41split} and match it with \eqref{eq:section-MP}. To obtain also $\C_0$, we use \eqref{eq:rest-weierstrass}. The result of the identification is
\begin{eqnarray}\label{eq:coeff-4-1-ng}
 \B  &=& c_1,  \nonumber\\
\C_0 &=& \tfrac14\,w^2\, (d_3^2 + 4\,w\,\alpha),  \nonumber\\
\C_1 &=&  \tfrac12\,w\,(-c_1 \, d_3 \, d_4 + 2\, w\, d_2), \\
\C_2 &=&  \tfrac14\, (c_1^2 \, d_4^2  + 4\, w\, c_0 \, d_4- 2\, w \, c_1 \, d_3), \nonumber\\
\C_3 &=&   w\,c_0+\tfrac12\,c_1^2 \, d_4 \nonumber\,.
\end{eqnarray}

Along the same lines one can also match the coefficients of the $3-2$ factorisation. The only difference to the above case is that special care is required in identifying the section because in \ref{sec:32split} the section is given in terms of the (torus) sum of two points. Taking this into account one finds
\begin{eqnarray}\label{eq:coeff-3-2-ng}
 \B  &=& \delta, \nonumber\\
\C_0 &=& \tfrac14\, w^2 \,(d_3^2\, \alpha^2 + 4\, w \,\alpha \,\gamma), \nonumber\\
\C_1 &=& \tfrac12\, w\, (c_2\, d_3^2\, \alpha + 2\,w\, ( d_2 \,\alpha + c_2\, \gamma)),\\
\C_2 &=& \tfrac14\, (c_2^2\, d_3^2 + 4\, w\, c_2\, d_2 - 2\,w\, d_3\,  \alpha\, \delta), \nonumber\\
\C_3 &=& w\, \beta - \tfrac12\,c_2\, d_3\, \delta \nonumber\,.
\end{eqnarray}

Indeed in both cases the base polynomials $\B, \C_i$ are of a non-generic form. In particular the powers of $w$ are responsible for the $SU(5)$ singularity along $w=0$.

\subsection{The \texorpdfstring{$SU(5)$}{SU(5)} resolution as a complete intersection}
The hypersurface equation \eqref{eq:211-hse-ng} with the coefficients \eqref{eq:coeff-4-1-ng} or\ \eqref{eq:coeff-3-2-ng} still exhibits an $SU(5)$-singularity at $w=0$. To resolve this singularity we find it more convenient to rewrite~\eqref{eq:211-hse-ng} in a form where all monomials in the homogeneous coordinates of $\textmd{Bl}_{[0,1,0]}\mathbb P_{[1,1,2]}$ with bi-degree $(4,3)$ show up. Equation~\eqref{eq:211-hse-ng} then becomes 
\begin{equation}\label{eq:211-hse-g}
\begin{split}
   B_2\,V^2\, W +  s\, W^2  + B_1\, s\, W\, V\, U  +&  B_0\,s^2\, W\, U^2 = C_3\,V^3\, U +\\+& C_2 s \, V^2 \, U^2 +  C_1 \, s^2 \,V\, U^3 +  C_0\, s^3\, U^4\,,
\end{split}
\end{equation}
where the coefficients for the $4-1$ factorisation are given by
\begin{eqnarray}\label{eq:coeff-4-1-g}
B_0  &=& -w\, d_3=w\,B_{0,1},  \nonumber\\
B_1  &=& c_1\, d_4=B_{1,0}, \nonumber\\
B_2  &=& c_1  = B_{2,0},\nonumber\\
C_0 &=& w^3\, \alpha  =w^3\, C_{0,3},\\
C_1 &=&  w^2 \, d_2 = w^2\,C_{1,2}, \nonumber\\
C_2 &=&  w\,c_0\, d_4  = w\,C_{2,1}, \nonumber\\
C_3 &=& w\,c_0   = w\,C_{3,1} \nonumber
\end{eqnarray}
 and for the  $3-2$ factorisation by
\begin{eqnarray}\label{eq:coeff-3-2-g}
B_0 &=&    -w\,d_3 \, \alpha = w\,B_{0,1},  \nonumber\\
B_1 &=&    -c_2\, d_3    = B_{1,0}, \nonumber\\
B_2 &=&    \delta    = B_{2,0},\nonumber\\
C_0 &=&   w^3\,\alpha\,\gamma = w^3\,C_{0,3},\\
C_1 &=&  w^2\,(d_2 \alpha + c_2 \gamma) = w^2\,C_{1,2}, \nonumber\\
C_2 &=&   w\,c_2\, d_2 = w\,C_{2,1}, \nonumber\\
C_3 &=&   w\,\beta = w\,C_{3,1}.\nonumber
\end{eqnarray}
To get back to \eqref{eq:211-hse-ng}, one just has to complete the square on the left-hand side of \eqref{eq:211-hse-g} and do a coordinate redefinition.

The fibration (\ref{eq:211-hse-g}) lends itself to a toric resolution of the singularity. 
From the classification of tops \cite{Bouchard:2003bu} one finds that for generic $B_{i,j}$ and $C_{i,j}$ there would be only an $SU(4)$ singularity at $W=V=w=0$. We start with the resolution of this $SU(4)$ singularity which is in both cases,  \eqref{eq:coeff-4-1-g} and \eqref{eq:coeff-3-2-g}, the same. Using an approach similar to \cite{Krause:2011xj}, which is actually equivalent to the top constructions of~\cite{Bouchard:2003bu}, we find the ambient five-fold $X_5$ of table~\ref{tab:ambient_fibre_space}
\begin{table}
 \begin{center}
  \begin{tabular}{c||c c c c |c c c |c || c}
             &  $V$      &  $W$      &   $U$       &   $s$       &   $e_1$     &   $e_2$     &   $e_3$   &   $e_0$    &  $P_T$   \\
   \hline
   \hline
   $[w]$     &  $\cdot$  &  $\cdot$  &   $\cdot$   &   $\cdot$   &   $\cdot$   &   $\cdot$   &   $\cdot$   &   $1$      &  $\cdot$ \\
   $\aK$     &  $1$      &  $2$      &   $\cdot$   &   $\cdot$   &   $\cdot$   &   $\cdot$   &   $\cdot$   &   $\cdot$  &  $4$\\
   $[\B]$      & $\cdot$   &  $\cdot$  &   $\cdot$   &   $1$       &   $\cdot$   &   $\cdot$   &   $\cdot$   &   $\cdot$  &  $1$     \\
   \hline
   $[U]$     &  $1$      &  $2$      &   $1$       &   $\cdot$   &   $\cdot$   &   $\cdot$   &   $\cdot$   &   $\cdot$  &  $4$     \\
   $[s]$     &  $1$      &  $1$      &   $\cdot$   &   $1$       &   $\cdot$   &   $\cdot$   &   $\cdot$   &   $\cdot$  &  $3$    \\
   \hline
   $E_1$     &  $\cdot$  &  $-1$     &   $\cdot$   &   $\cdot$   &   $1$       &   $\cdot$   &   $\cdot$   &   $-1$     &  $-1$    \\
   $E_2$     &  $-1$     &  $-2$     &   $\cdot$   &   $\cdot$   &   $\cdot$   &   $1$       &   $\cdot$   &   $-1$     &  $-3$    \\
   $E_3$     &  $-1$     &  $-1$     &   $\cdot$   &   $\cdot$   &   $\cdot$   &   $\cdot$   &   $1$       &   $-1$     &  $-2$    \\
   \hline
   \hline
             &  $ 1$     &  $0$      &   $-1$      &   $-1$      &   $0$       &   $1$       &   $1$       &   $0$ &     \\
             &  $-1$     &  $1$      &   $-1$      &   $0$       &   $1$       &   $1$       &   $0$       &   $0$ &    \\
             &  $\underline{0}$   &   $\underline{0}$   &   $\underline{0}$   &   $\underline{0}$   &   $\underline{v}$   &   $\underline{v}$   &   $\underline{v}$   &   $\underline{v}$
  \end{tabular}
 \caption{Divisor classes and coordinates of the ambient space with $V$, $W$, $U$, $s$ the coordinates of the ``fibre ambient space" of the Calabi-Yau four-fold. 
Note that the base classes $W$, $\B$ and $\aK = c_1(B)$ are included.
The bottom of the table is only relevant if the entire 4-fold including the base is torically embedded. It lists a choice for the vectors corresponding to the one-cones of the toric fan.}
\label{tab:ambient_fibre_space}
 \end{center}
\end{table}
and the proper transform of the hypersurface equation taking the form
\begin{equation}\label{eq:211-hse-su4}
 \begin{split}
  e_2 e_3\, B_{2,0}\,V^2\, W &+ e_1 e_2\, s\, W^2  + B_{1,0}\, s\, W\, V\, U  +  e_1\, e_0\, B_{0,1}\,s^2\, W\, U^2 =e_2 e_3^2\,  e_0\, C_{3,1}\,V^3\, U +\\
&+  e_3\, e_0 \, C_{2,1} s \, V^2 \, U^2 +  e_1 e_3\,e_0^2 \, C_{1,2} \, s^2 \,V\, U^3 +   e_1^2 e_3\,e_0^3 \, C_{0,3}\, s^3\, U^4.
\end{split}
\end{equation}
This resolution allows for different Stanley-Reisner ideals. For brevity we use the following Stanley-Reisner ideal in the sequel, 
\begin{equation}\label{eq:211-sri}
\{V\,s,\, V\,e_1,\, W\,U,\, W\,e_0,\, W\,e_3,\, U\,e_2,\, s\,e_2,\, e_0\,e_2,\, U\,e_3,\, s\,e_0,\, s\,e_3 \}\,.
\end{equation}
Note that (\ref{eq:211-hse-su4}) (with generic coefficients) describes only one out of several tops realising an $SU(4)$ singularity over $w=0$ in a $\textmd{Bl}_{[0,1,0]}\mathbb P_{[1,1,2]}$-fibration.  In \cite{inprogress} we will discuss  the remaining options and provide a more detailed description of the resolution procedure.

Due to the non-genericity of the coefficients, equation \eqref{eq:211-hse-su4} is still singular. This follows from the fact that it factorises, concretely for the $4-1$ model as
\begin{equation}
 \label{eq:211-hse-su4-4-1}
e_1\,s\,Q = V\,P_1\,P_2 
\end{equation}
with
\begin{eqnarray}
         Q &=& -e_2\, W^2 + e_0\,d_3 \, s\, W\, U^2 +   e_3\, e_0^2\,d_2\,s\, V\, U^3 +    e_1 e_3\,e_0^3\, \alpha\, s^2\,U^4, \nonumber\\
       P_1 &=& e_2 e_3\, V+ d_4\, s\, U,\\
       P_2 &=& c_1\, W - e_3\,e_0\, c_0 \, V\,U\nonumber
\end{eqnarray}
and for the $3-2$ model as
\begin{equation}
 \label{eq:211-hse-su4-3-2}
e_2\,\Q = s\,U\,\PP_1\,\PP_2
\end{equation}
with
\begin{eqnarray}
  \Q &=& e_1\, s\, W^2 - e_3^2\,e_0 \, \beta\, V^3\, U  + e_3 \,\delta\,V^2\, W, \nonumber\\
\PP_1 &=&  e_3\, e_0\,d_2\, U\,V + d_3\, W +  e_1 e_3\,e_0^2\,\gamma\,s\,U^2, \\
\PP_2 &=& c_2\, V + e_1\, e_0\,\alpha\, s\, U.\nonumber
\end{eqnarray} 
From \eqref{eq:211-hse-su4-4-1} and \eqref{eq:211-hse-su4-3-2} and the Stanley-Reisner ideal one easily observes that the exceptional divisors $E_1= \{e_1 =0 \}$ and $E_2= \{ e_2 =0 \}$, respectively, split into two on the hypersurface. Since the two parts of the factorised divisor intersect each other, we obtain the $\pl{}$-structure of an $SU(5)$-singularity.

Due to the Stanley-Reisner ideal~\eqref{eq:211-sri} only $e_1$, $Q$, $P_1$ and $P_2$, for the $4-1$ model, and  $e_2$, $\Q$, $\PP_1$ and $\PP_2$, for the $3-2$ model, have a common locus. Therefore, the last resolution step is in both cases a small resolution given, respectively, by
\begin{equation}\label{eq:211-cicy-4-1}
 \lambda_1\, e_1\,s= \lambda_2\, P_2\,,\qquad  \lambda_2\, Q= \lambda_1\, V\,P_1
\end{equation}
and
\begin{equation}\label{eq:211-cicy-3-2}
  \lambda_1\, e_2 = \lambda_2\,s\,\PP_2\,,\qquad  \lambda_2\,\Q= \lambda_1\, U\,\PP_1\,,
\end{equation}
with $\lambda_1$ and $\lambda_2$ the homogeneous coordinates of some appropriate line bundle over the blown-up ambient space $X_5$. Therefore, we obtain in both cases an ambient six-fold, $X_6$ and $\textmd X_6$, in which the four-fold is described as a complete intersection.

Inspection of the gradient to the hypersurface shows that the only possible remaining singularities in this procedure are at the intersection of $\gamma=\alpha=0$.
There are two ways to nonetheless achieve a well-defined fourfold: Either one restricts oneself to fibrations over base spaces $B_3$ where the intersection structure excludes this locus. Alternatively one can further restrict the fibration such that  $\gamma$ or $\alpha$ are constant, i.e. transform as sections of the trivial bundle. Indeed, in the appendix of \cite{inprogress} we show that exploiting the freedom in the definition of the base sections $\alpha, \beta, \gamma, \delta, d_2, d_3$ one can set $\alpha$ constant while maintaining two ${\bf 10}$-curves. This procedure leads to a well-defined complete intersection provided the remaining base sections $\beta, \ldots, d_3$ can be chosen as sections of positive line bundles on $B_3$. As shown in \cite{inprogress}  this condition is fulfilled e.g. for $B_3=\mathbb P^3$ \cite{inprogress}, thereby providing a well-defined elliptic fibration defined as a complete intersection with $SU(5)$ gauge group and two ${\bf 10}$-curves.

We can now examine the $\mathbb P^1$-structure in the resolved geometry. We start in co-dimension one with $w=0$.
Here, as we mentioned already, the only difference to the `standard' $SU(5)$ case is that two of the $\pl{}$'s in the fibre come from the splitting of one of the $\pl{}$'s of the $SU(4)$ resolution. The fibration of these $\mathbb P^1$s over $w=0$ give rise to a set of divisors $\E_i$ intersecting like the Cartan of $SU(5)$ if we adopt the respective labellings
\begin{equation}
\begin{split}
 \E_0 &= [e_0 \cap P_\textmd{CI}],   \\
 \E_1 &= [e_1 \cap \,P_1\cap    \lambda_2],     \\
 \E_2 &= [e_1 \cap P_2   \cap   \lambda_2\, Q- \lambda_1\,P_1 ],       \\
 \E_3 &= [e_2 \cap P_\textmd{CI}   ],\\
 \E_4 &= [e_3 \cap P_\textmd{CI} ],
\end{split}
\qquad\textmd{and}\qquad
\begin{split}
 \E_0 &= [e_0 \cap \PP_\textmd{CI} ],  \\
 \E_1 &= [e_1 \cap \PP_\textmd{CI}  ], \\
 \E_2 &= [e_2 \cap \PP_1 \cap    \lambda_2],      \\
 \E_3 &= [e_2 \cap \PP_2 \cap \lambda_2\, \Q- \lambda_1\,\PP_1 ],     \\
 \E_4 &= [e_3 \cap \PP_\textmd{CI} ],
\end{split}
\end{equation}
where $P_\textmd{CI}$ and $\PP_\textmd{CI}$ refer to the complete intersection~\eqref{eq:211-cicy-4-1} and \eqref{eq:211-cicy-3-2}, respectively.

\subsection{The singlet curves in the resolved \texorpdfstring{$\mathbb P_{[1,1,2]}$}{P[1,1,2]}-fibration}
It is straightforward to re-analyse the structure of the fibres over the matter curves and the Yukawa interaction points starting from the resolved 4-folds (\ref{eq:211-cicy-4-1}) and (\ref{eq:211-cicy-3-2}). We do not spell out the details of this analysis here, but merely note that the findings of sections \ref{sec:fibstruc-gen} and \ref{sec:fibu1pq} are indeed confirmed except for slight details in the structure of the Yukawa points.\footnote{These, however, change for different triangulations anyway.} Concerning the analysis of the co-dimension three singularities we point out that at the $\overline{\mathbf 5}^{(1)}_{-2} \mathbf 5^{(2)}_{-3} \mathbf 1_{5}$ point we find a fibre of extended $A_6$ type.

What we do present now is an analysis of the charged singlet curves, which as discussed in section  \ref{sec:fibu1pq} is subtle in the small resolution approach of the previous sections. Indeed, the structure of $U(1)$ charged singlets in the resolved $\mathbb P_{[1,1,2]}$ model (\ref{eq:211-hse-g}) has been worked out in detail in \cite{Morrison:2012ei} for a fibration over the base $\mathbb P^2$. Since the appearance and further resolution of the $SU(5)$ singularity over $w=0$ is irrelevant for the generic points on the singlet curves, we can adopt this analysis.
According to the general logic of  \cite{Morrison:2012ei}, it then follows  that the singlets of $U(1)$ charge $\pm 10$ localise in the fibre over the curve
\begin{equation}\label{eq:10-locus-generic}
 \B=\C_3=0\,.
\end{equation}
The charge $\pm5$ singlets are located at the \emph{generic} intersection of the two loci
\begin{equation}
 \begin{split}\label{eq:5-locus-generic}
  0 & = -\tfrac12\, B^4\, \C_1 + B^2\,\C_2\,\C_3 - \C_3^3, \\
  0 & = -B^6\, \C_0 + B^4\, \C_2^2 - 2\, B^2\, \C_2\, \C_3^2 + \C_3^4 \,
 \end{split}
\end{equation}
where $\B$ and $\C_3$ do not simultaneously vanish.

To read off the charges, we  recall that these were given by the intersection of the $\pl{}$'s of the resolved singularity with the divisor
\[\textmd{w}_X=5(S-Z \ldots),\]
where the ellipsis indicates terms irrelevant for the current explanation. In \cite{Morrison:2012ei} it was now shown that at the locus \eqref{eq:10-locus-generic} equation 
\eqref{eq:211-hse-ng} becomes
\[s\,D=0\,,\]
whereas at \eqref{eq:5-locus-generic} it factorises as
\[ (A-B)(A+B)=0\,,\]
where $A$, $B$ and $D$ are some polynomials. Therefore, the two $\pl{}$'s into which the torus factorises are in the one case
\[\pl{s}:\quad s=0\qquad\textmd{and}\qquad \pl{D}:\quad D=0\]
and in the other case
\[\pl{A-}:\quad A-B=0\qquad \textmd{and}\qquad \pl{A+}:\quad A+B=0.\]
To calculate now the charge for $\pl{D}$ we observe from figure~\ref{fig:U1-sing}
\begin{figure} 
 \centering
 \def\svgwidth{0.65\linewidth}
 \executeiffilenewer{U1-sing.svg}{U1-sing.pdf}%
 {inkscape -z -D --file=U1-sing.svg %
  --export-pdf=U1-sing.pdf --export-latex}%
   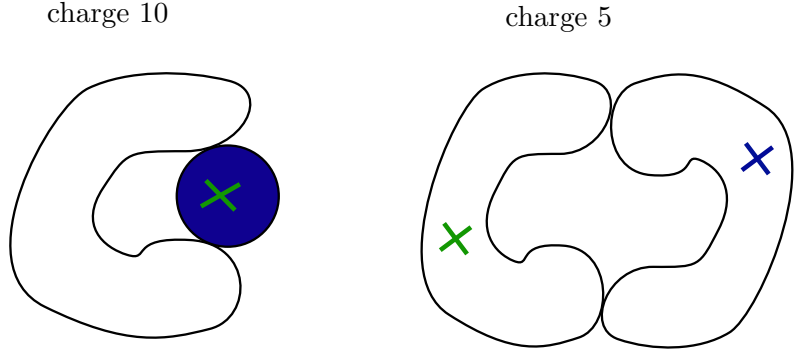%

 \caption{Schematic drawing of the fibre over the charge 10 and 5 singlet curves. The green and blue crosses indicate the intersections with $Z$ and $S$, respectively. In the case of the charge $10$ singlet, $S$ becomes one of the $\pl{}$'s, which we indicate by the blue ball.}
\label{fig:U1-sing}
\end{figure}
that it intersects $S$ two times but does not have any overlap with $Z$, since $Z$ intersects only $\pl{s}$. Hence, we obtain $+10$ for $\int_{ \pl{D}} \textmd{w}_X$. As was already explained in the discussion of the singlets in sections~\ref{sec:fibstruc-gen} and \ref{sec:fibu1pq}, the M2-brane wrapping the second $\pl{}$ is just the adjoint state to the M2-brane wrapping the other one. Therefore, $\pl{s}$ must have charge $-10$. From figure~\ref{fig:U1-sing} it is also clear that $\int_{\pl{A-}} \textmd{w}_X =- \int_{\pl{A+}}\textmd{w}_X=5$. Again M2-branes wrapping $\pl{A-}$ and $\pl{A+}$ are  adjoint states to each other.

In the $4-1$ case, a simultaneous vanishing of $c_1$ and $c_0$ is forbidden because this would lead to non-Kodaira singularities as stated after eq.\ (\ref{d0d1X}). Therefore, there are no singlets ${\bf 1}_{\pm 10}$ but only ${\bf 1}_{\pm 5}$ states localised on the curve
\begin{equation}
 \begin{split}
  0 &  =-\tfrac12 w^2 (c_1^2\,(c_1^2\, d_2 + c_0\, c_1\, d_3 + c_0^2\, d_4) + 2\, c_0^3\, w),\\
  0 &  = w^3\, (c_0^2\,( c_1^3\, d_3 + c_0^2\, w) + c_1^6\, \alpha),
 \end{split}
\end{equation}
in agreement with the results of section {\ref{sec:fibstruc-gen}.
For the $3-2$ model the situation is different because now we also have charge $\pm 10$ singlets at
\begin{equation}
 \delta=\beta=0\,,
\end{equation}
besides the charge five singlets at
\begin{equation}
\begin{split}
0 & = -\tfrac12 \,w\, \Big(2\, w^2\,\beta^3 + c_2^2\, d_3\,\delta^2\, (d_3 \,\beta + d_2\, \delta) + w\,\delta \,\big(\alpha\,\delta^2 \,(d_3\, \beta + d_2 \,\delta)+\\
  &\qquad+ c_2\, (\gamma\, \delta^3-3\, d_3\, \beta^2 - 2\, d_2\, \beta\, \delta )\big)\Big)\\
0 & = w^2\, \Big(w^2 \,\beta^4 + c_2\, \delta^2\, (d_3 \,\beta + d_2\, \delta)\, (c_2\, d_3\, \beta + c_2\, d_2\, \delta - d_3 \,\alpha\, \delta^2) +\\
  &\qquad- w\, \delta\, \big( \alpha\, \gamma\, \delta^5-d_3\, \alpha\, \beta^2\, \delta^2  + 2\, c_2\, \beta^2\, (d_3\, \beta + d_2\, \delta)\big)\Big)
 \end{split}
\end{equation}
and $\delta\ne0$ or $\beta\ne0$.

Note again that we are explicitly excluding the locus $\alpha = \gamma = 0$, where potential singularities may remain---see the discussion after (\ref{eq:211-cicy-3-2}) and \cite{inprogress}.

\section{Summary}
\label{sec:disc} 

In this article we have studied 4-dimensional F-theory compactifications with $U(1)$ symmetries in addition to a non-Abelian gauge group $G$, taken to be $SU(5)$ for definiteness. We developed a systematic approach to construct such models as factorised Tate models with multiple sections. We have provided the form of the factorised Tate models for a wide range of possible $U(1)$ symmetries and exemplified the resolution of the associated singular fibrations for the two cases with a single $U(1)$  - called $SU(5) \times U(1)_X$ and $SU(5) \times U(1)_{PQ}$.
These can be treated either patchwise by a small resolution procedure or in terms of the resolution of a $\mathbb P_{[1,1,2]}$ fibration of the type introduced, without further non-Abelian gauge symmetry, in \cite{Morrison:2012ei}. Our resolution of the $SU(5)$ singularities of the  $\mathbb P_{[1,1,2]}$ fibration has been achieved by supplementing a toric blow-up procedure by a small resolution. This leads to a well-defined four-fold described as a complete intersection in an ambient six-fold.
Our results obtained in both approaches concerning the fibre structure agree. 
In particular the fibrations studied in this work provide the first examples of F-theory $SU(5)$ compactifications with two ${\bf 10}$-curves, which are of current phenomenological interest, e.g.\ in the context of flavour physics \cite{Dudas:2009hu}.

An explicit construction of the $U(1)$ generators after the resolution and analysis of the intersection with the fibres over the matter curves has allowed us to derive the Abelian charges of all matter states directly from the geometry. As in \cite{Grimm:2010ez,Braun:2011zm,Krause:2011xj,Grimm:2011fx} this also provides us with the associated $U(1)$ flux for chiral model building.

The $SU(5) \times U(1)_X$ model is a generalisation of the $U(1)$ restricted Tate model \cite{Grimm:2010ez,Braun:2011zm,Krause:2011xj,Grimm:2011fx} and all the factorised Tate models flow, in the vicinity of the $SU(5)$ divisor, to the split spectral models \cite{Marsano:2009gv,Marsano:2009wr,Dudas:2010zb,Dolan:2011iu}.
What makes an analysis of Abelian gauge groups within a fully global treatment of the geometry so crucial is the fact that $U(1)$ symmetries are sensitive to geometric details away from the $SU(5)$ divisor. In particular, charged singlets arise from curves extending into the bulk of the compactification space, where local methods fail. 
As one of the surprises we have encountered, in the global version of the $SU(5) \times U(1)_{PQ}$ model an extra set of charged singlets of charge $\pm 10$ appears, which apparently do not follow from a decomposition of the ${\bf 248}$ of a single $E_8$ gauge group. Since these states couple to the $SU(5)$ matter these novel states can in principle influence the phenomenology of the model. 
The extension of our methods to a detailed analysis of the models with several Abelian factors as classified in this paper is under way \cite{inprogress}.

We hope that the systematic construction of global F-theory GUT models exhibiting additional $U(1)$ symmetries, without restricting the matter spectrum or Yukawa couplings, outlined in this paper will open the way to realising much of the successful phenomenology of local F-theory models in a fully global setting.

\subsection*{Acknowledgements}

We thank Max Kerstan for initial collaboration and Harald Skarke  for very useful discussions. We also thank Jan Borchmann for providing the details of the $A_3$ top and Kang-Sin Choi for discussion. 
CM thanks IPhT CEA Saclay, especially Mariana Gra\~na,  for hospitality.
Furthermore, we are grateful to David Morrison and Daniel Park for providing us with a draft of their paper.
The research of EP is supported by a Marie Curie  Intra European Fellowship within the 7th European Community Framework Programme.
The work of CM and TW was funded in part by the DFG under Transregio TR 33 "The Dark Universe".

\appendix

\section{More factorised Tate models}
\label{sec:morfactate}

In this appendix we give the form of (\ref{tatesing}) explicitly for the other possible factorisations. We use here the notation of \cite{Dudas:2010zb} and in particular denote products $c_i...c_j=c_{i...j}$.

\subsection{\texorpdfstring{$2 - 2 - 1$}{2 - 2 - 1} Factorisation}
\label{sec:221split}

The factorisation of $\left.P_T\right|_{X=0}$ before the $SU(5)$ resolution is
\begin{equation}
\left(c_1 t^2  + c_2 t  + c_3 \right)\left( c_4 t^2  +  c_5 t  + c_6\right)\left(c_7 t  + c_8 \right) \;.
\end{equation} 
The $a_{i,n}$ are given by
\begin{eqnarray}
a_{6,5} &=& c_{368} \;, \nn \\
a_{5,4} &=& c_{367} + c_{358} + c_{268}\;, \nn \\
a_{4,3} &=& c_{357} + c_{267} + c_{348} + c_{258} + c_{168}\;, \nn \\
a_{3,2} &=& c_{347} + c_{257} + c_{167} + c_{248} + c_{158}\;, \nn \\
a_{2,1} &=& c_{247} + c_{157} + c_{148} \;, \nn \\
a_{1} &=& c_{147} \;. \label{aIsol221}
\end{eqnarray}
A solution to the tracelessness constraint was given in \cite{Dolan:2011iu} and reads
\begin{eqnarray}
c_3 &=& \alpha \beta \delta_1 \;, \nn \\
c_2 &=& \gamma \delta_1 \;, \nn \\
c_6 &=& \alpha \beta \epsilon \delta_2 \;, \nn \\
c_5 &=& - \delta_2 \left(\gamma \epsilon + c_7 \beta \right) \;, \nn \\
c_8 &=& \alpha \epsilon \;. \label{b1sol221}
\end{eqnarray}
Unlike the case of a single $U(1)$ there now appears a subtlety in defining this solution because $a_{5,4}$ is a sum of 3 terms but can be set to vanish with only 2 sections. So for example setting $\epsilon=\beta=0$ solves $a_{5,4}=0$ without imposing $c_3=0$ necessarily, even though the solution (\ref{b1sol221}) would constrain it to be so. There are a number of such special cases that occur on the intersection locus of 2 sections. Therefore placing constraints on the intersection numbers in order to avoid non-Kodaira singularities is very complicated, since the solution (\ref{b1sol221}) could be adjusted accordingly on these special loci to avoid such a bad singularity. Hence the solution (\ref{b1sol221}) is not the most general one. In this paper we will not perform an analysis of the most general solution and the resulting constraints on intersection numbers to avoid non-Kodaira singularities. We will work explicitly with the solution (\ref{b1sol221}) and leave the most general analysis for future work.

With this solution, in order to ensure no non-Kodaira singularities, for generic sections we should impose that the following intersections must vanish
\begin{equation}
\alpha\cdot c_7 \;,\; \epsilon\cdot  c_7  \;,\; c_1\cdot  \delta \;,\; c_4\cdot  \delta \;. 
\end{equation} 
Since the sections are now objects over the whole base we should also consider triple intersections which must vanish
\begin{equation}
c_1 \cdot \alpha \cdot \gamma \;,\; c_1 \cdot \beta \cdot \gamma \;,\; c_4 \cdot \gamma \cdot \epsilon \;,\; c_4 \cdot c_7 \cdot \beta \;.
\end{equation} 

Within the patch $e_1=e_2=e_4=1$ we have that the Tate polynomial, after the $SU(5)$ resolution, can be written as (\ref{tatesing}) with 
\begin{eqnarray}
Y_1 &=& c_7 t + \alpha e_0 \epsilon z\;, \nn \\
Y_2 &=& c_1 t^2 + \delta_1 e_0 \gamma t z + \alpha \beta \delta_1 e_0^2 z^2\;, \nn \\
Y_3 &=& c_4 t^2 - \beta c_7 \delta_2 e_0 t z - \delta_2 e_0 \epsilon \gamma t z + \alpha \beta \delta_2 e_0^2 \epsilon z^2\;, \nn \\
X &=& t^2 - x \;, \nn \\
Q &=& e_3 x^2 + c_1 c_4 c_7 t^3 z + c_1 c_4 c_7 t x z - \beta c_1 c_7^2 \delta_2 e_0 t^2 z^2 + 
  \alpha c_1 c_4 e_0 \epsilon t^2 z^2 + c_4 c_7 \delta_1 e_0 \gamma t^2 z^2 - 
  c_1 c_7 \delta_2 e_0 \epsilon \gamma t^2 z^2 \nn \\
  & &- \beta c_1 c_7^2 \delta_2 e_0 x z^2 + 
  \alpha c_1 c_4 e_0 \epsilon x z^2 + c_4 c_7 \delta_1 e_0 \gamma x z^2 - 
  c_1 c_7 \delta_2 e_0 \epsilon \gamma x z^2 + \alpha \beta c_4 c_7 \delta_1 e_0^2 t z^3 - 
  \beta c_7^2 \delta_1 \delta_2 e_0^2 \gamma t z^3 \nn \\
  & &+ \alpha c_4 \delta_1 e_0^2 \epsilon \gamma t z^3 - 
  \alpha c_1 \delta_2 e_0^2 \epsilon^2 \gamma t z^3 - c_7 \delta_1\delta_2 e_0^2 \epsilon \gamma^2 t z^3 - 
  \alpha \beta^2 c_7^2 \delta_1\delta_2 e_0^3 z^4 + \alpha^2 \beta c_4 \delta_1 e_0^3 \epsilon z^4 + 
  \alpha^2 \beta c_1 \delta_2 e_0^3 \epsilon^2 z^4 \nn \\ & &- \alpha \beta c_7 \delta_1\delta_2 e_0^3 \epsilon \gamma z^4 - 
  \alpha \delta_1\delta_2 e_0^3 \epsilon^2 \gamma^2 z^4 \;.
\end{eqnarray}

In the case of multiple $U(1)$s the single small resolution (\ref{compint}) of course is not sufficient to completely resolve the manifold but a generalisation of it is required. The resolution of this particular type of binomial singularity was studied in detail in \cite{Esole:2011sm}. We introduce two new $\mathbb P^1$s spanned by $\left\{\lambda_1,\lambda_2\right\}$ and $\left\{\sigma_1,\sigma_2\right\}$, in terms of which the resolved four-fold is given by
\begin{eqnarray}
\tilde{Y}_4\;:\;\left\{Y_1 \lambda_2 = Q \lambda_1 \right\} \cap \left\{Y_2 \sigma_2 = X \sigma_1 \right\} \cap \left\{\lambda_2 \sigma_2 = \lambda_1 \sigma_1 Y_3 \right\} \;. \label{resy2u1}
\end{eqnarray}
There are 6 different possible resolutions, related by flop transitions, corresponding to permuting $\left\{Y_1,Y_2,Y_3\right\}$ in (\ref{resy2u1}).

\subsection{\texorpdfstring{$3 - 1 - 1$}{3 - 1 - 1} Factorisation}
\label{sec:311split}

In this case we have the factorisation of $\left.P_T\right|_{X=0}$ as
\begin{equation}
\left(c_1 t^3  + c_2 t^2 +c_3 t + c_4 \right)\left(c_5 t + c_6 \right)\left(c_7 t + c_8\right) \;.
\end{equation} 
The $a_{i,n}$ are given by
\begin{eqnarray}
a_{6,5} &=& c_{468} \;, \nn \\
a_{5,4} &=& c_{467} + c_{458} + c_{368}\;, \nn \\
a_{4,3} &=& c_{457} + c_{367} + c_{358} + c_{268}\;, \nn \\
a_{3,2} &=& c_{357} + c_{267} + c_{258} + c_{168}\;, \nn \\
a_{2,1} &=& c_{257} + c_{167} + c_{158} \;, \nn \\
a_{1} &=& c_{157} \;. \label{aIsol311}
\end{eqnarray}
A possible solution to the tracelessness constraint was given in \cite{Dolan:2011iu} and reads
\begin{eqnarray}
c_6 &=& \alpha \beta \;, \nn \\
c_8 &=& \alpha \gamma \;, \nn \\
c_4 &=& \alpha \beta \gamma \delta \;, \nn \\
c_3 &=& -\delta \left( c_5 \gamma + \beta c_7 \right)\;. \label{b1sol311}
\end{eqnarray}
With this solution, in order to ensure no non-Kodaira singularities, for generic sections we should impose that the following intersections must vanish
\begin{eqnarray}
c_5 \cdot \alpha \;,\; c_7 \cdot \alpha \;,\; c_5 \cdot \beta \;,\; c_7 \cdot \gamma \;,\; c_1 \cdot c_2 \cdot \delta \;.
\end{eqnarray}
Note that again, as discussed in section \ref{sec:221split}, the solution (\ref{b1sol311}) is not the most general one.

Within the patch $e_1=e_2=e_4=1$ we have that the Tate polynomial can be written as (\ref{tatesing}) with 
\begin{eqnarray}
Y_1 &=& c_5 t + \alpha \beta e_0 z\;, \nn \\
Y_2 &=& c_7 t + \alpha e_0 \gamma z\;, \nn \\
Y_3 &=& c_1 t^3 + c_2 e_0 t^2 z - \beta c_7 \delta e_0^2 t z^2 - c_5 \delta e_0^2 \gamma t z^2 + 
   \alpha \beta \delta e_0^3 \gamma z^3\;, \nn \\
X &=& t^2 - x \;, \nn \\
Q &=& e_3 x^2 + c_1 c_5 c_7 t^3 z + c_1 c_5 c_7 t x z + \alpha \beta c_1 c_7 e_0 t^2 z^2 + 
 c_2 c_5 c_7 e_0 t^2 z^2 + \alpha c_1 c_5 e_0 \gamma t^2 z^2 + \alpha \beta c_1 c_7 e_0 x z^2 \nn \\
  & & +c_2 c_5 c_7 e_0 x z^2 + \alpha c_1 c_5 e_0 \gamma x z^2 + \alpha \beta c_2 c_7 e_0^2 t z^3 - 
 \beta c_5 c_7^2 \delta e_0^2 t z^3 + \alpha^2 \beta c_1 e_0^2 \gamma t z^3 + 
 \alpha c_2 c_5 e_0^2 \gamma t z^3 \nn \\
  & &-c_5^2 c_7 \delta e_0^2 \gamma t z^3 - \alpha \beta^2 c_7^2 \delta e_0^3 z^4 + 
 \alpha^2 \beta c_2 e_0^3 \gamma z^4 - \alpha \beta c_5 c_7 \delta e_0^3 \gamma z^4 - 
 \alpha c_5^2 \delta e_0^3 \gamma^2 z^4 \;.
\end{eqnarray}

\subsection{\texorpdfstring{$2 - 1 - 1 - 1$}{2 - 1 - 1 - 1} Factorisation}
\label{sec:2111split}

The factorisation is
\begin{equation}
\left(c_1 t^2 + c_2 t + c_3 \right) \left(c_4 t + c_7 \right) \left(c_5 t + c_8 \right) \left(c_6 t + c_9 \right) \;. \label{spect10z2111}
\end{equation} 
The $a_{i,n}$ are given by
\begin{eqnarray}
a_{6,5} &=& c_{3789} \;, \nn \\
a_{5,4} &=& c_{2789} + c_{3678} + c_{3579} + c_{3489} \;, \nn \\
a_{4,3} &=& c_{1789} + c_{2678} + c_{2579} + c_{2489} + c_{3567} + c_{3468} + c_{3459} \;, \nn \\
a_{3,2} &=& c_{3456} + c_{1678} + c_{1579} + c_{1489} + c_{2567} + c_{2468} + c_{2459} \;, \nn \\
a_{2,1} &=& c_{2456} + c_{1567} + c_{1468} + c_{1459} \;, \nn \\
a_{1} &=& c_{1456} \;. \label{aIsol2111}
\end{eqnarray}
A solution to the tracelessness constraint is
\begin{eqnarray}
c_7 &=& \delta \alpha_1 \;, \nn \\
c_8 &=& \delta \alpha_2 \;, \nn \\
c_9 &=& \delta \alpha_3 \;, \nn \\
c_3 &=& \delta \epsilon \alpha_1 \alpha_2 \alpha_3 \;, \nn \\
c_2 &=& -\epsilon \left(c_6 \alpha_1\alpha_2 + c_5 \alpha_1\alpha_3 + c_4 \alpha_2\alpha_3\right)\;.
\end{eqnarray}
The intersections which must vanish are
\begin{eqnarray}
& &c_4 \cdot \delta \;,\; c_5 \cdot \delta \;,\; c_6 \cdot \delta \;,\; c_4 \cdot \alpha_1 \;,\; c_5 \cdot \alpha_2 \;,\; c_6 \cdot \alpha_3 \;,\; c_1 \cdot \epsilon \;, \nn \\
& &c_1 \cdot \alpha_1 \cdot \alpha_2 \;,\; c_1 \cdot \alpha_1 \cdot \alpha_3 \;,\; c_1 \cdot \alpha_2 \cdot \alpha_3 \;, \nn \\
& &c_1 \cdot c_6 \cdot \alpha_3 \;,\; c_1 \cdot c_5 \cdot \alpha_2 \;,\; c_1 \cdot c_4 \cdot \alpha_1 \;.
\end{eqnarray}
Within the patch $e_1=e_2=e_4=1$ we have that the Tate polynomial can be written as (\ref{tatesing}) with 
\begin{eqnarray}
Y_1 &=& c_4 t + \alpha_1 \delta e_0 z\;, \nn \\
Y_2 &=& c_5 t + \alpha_2 \delta e_0 z \;, \nn \\
Y_3 &=& c_6 t + \alpha_3 \delta e_0 z\;, \nn \\
Y_4 &=& c_1 t^2 - \alpha_2 \alpha_3 c_4 e_0 \epsilon t z - \alpha_1 \alpha_3 c_5 e_0 \epsilon t z - 
  \alpha_1 \alpha_2 c_6 e_0 \epsilon t z + \alpha_1 \alpha_2 \alpha_3 \delta e_0^2 \epsilon z^2\;, \nn \\
X &=& t^2 - x \;, \nn \\
Q &=&  e_3 x^2 + c_1 c_4 c_5 c_6 t^3 z + c_1 c_4 c_5 c_6 t x z + 
 \alpha_3 c_1 c_4 c_5 \delta e_0 t^2 z^2 + \alpha_2 c_1 c_4 c_6 \delta e_0 t^2 z^2 + 
 \alpha_1 c_1 c_5 c_6 \delta e_0 t^2 z^2 \nn \\
  & & - \alpha_2 \alpha_3 c_4^2 c_5 c_6 e_0 \epsilon t^2 z^2 - 
 \alpha_1 \alpha_3 c_4 c_5^2 c_6 e_0 \epsilon t^2 z^2 -\alpha_1 \alpha_2 c_4 c_5 c_6^2 e_0 \epsilon t^2 z^2 + \alpha_3 c_1 c_4 c_5 \delta e_0 x z^2 + 
 \alpha_2 c_1 c_4 c_6 \delta e_0 x z^2 \nn \\
  & & + \alpha_1 c_1 c_5 c_6 \delta e_0 x z^2 - 
 \alpha_2 \alpha_3 c_4^2 c_5 c_6 e_0 \epsilon x z^2 - \alpha_1 \alpha_3 c_4 c_5^2 c_6 e_0 \epsilon x z^2 - \alpha_1 \alpha_2 c_4 c_5 c_6^2 e_0 \epsilon x z^2 + \alpha_2 \alpha_3 c_1 c_4 \delta^2 e_0^2 t z^3 \nn \\
  & & + \alpha_1 \alpha_3 c_1 c_5 \delta^2 e_0^2 t z^3 + \alpha_1 \alpha_2 c_1 c_6 \delta^2 e_0^2 t z^3 - 
 \alpha_2 \alpha_3^2 c_4^2 c_5 \delta e_0^2 \epsilon t z^3 - 
 \alpha_1 \alpha_3^2 c_4 c_5^2 \delta e_0^2 \epsilon t z^3 -\alpha_2^2 \alpha_3 c_4^2 c_6 \delta e_0^2 \epsilon t z^3  \nn \\
  & &  - 2 \alpha_1 \alpha_2 \alpha_3 c_4 c_5 c_6 \delta e_0^2 \epsilon t z^3 - 
 \alpha_1^2 \alpha_3 c_5^2 c_6 \delta e_0^2 \epsilon t z^3 - 
 \alpha_1 \alpha_2^2 c_4 c_6^2 \delta e_0^2 \epsilon t z^3 - 
 \alpha_1^2 \alpha_2 c_5 c_6^2 \delta e_0^2 \epsilon t z^3 \nn \\ & &
 + \alpha_1 \alpha_2 \alpha_3 c_1 \delta^3 e_0^3 z^4
 -\alpha_2^2 \alpha_3^2 c_4^2 \delta^2 e_0^3 \epsilon z^4 - 
 \alpha_1 \alpha_2 \alpha_3^2 c_4 c_5 \delta^2 e_0^3 \epsilon z^4 - 
 \alpha_1^2 \alpha_3^2 c_5^2 \delta^2 e_0^3 \epsilon z^4 - 
 \nn \\ & &
 \alpha_1 \alpha_2^2 \alpha_3 c_4 c_6 \delta^2 e_0^3 \epsilon z^4 - 
 \alpha_1^2 \alpha_2 \alpha_3 c_5 c_6 \delta^2 e_0^3 \epsilon z^4 - \alpha_1^2 \alpha_2^2 c_6^2 \delta^2 e_0^3 \epsilon z^4 \;.
\end{eqnarray}

\subsection{\texorpdfstring{$1 - 1- 1 - 1 - 1$}{1 - 1- 1 - 1 - 1} Factorisation}
\label{sec:11111split}

The factorisation is
\begin{equation}
\left(c_1 t+ c_6 \right) \left(c_2 t+ c_7 \right) \left(c_3 t + c_8 \right) \left(c_4 t+ c_9 \right) \left(c_5 t+ c_{10} \right) \;. \label{spect10z11111}
\end{equation} 
The $a_{i,n}$ are given by
\begin{eqnarray}
a_{6,5} &=& c_{10} c_6 c_7 c_8 c_9 \;, \nn \\
a_{5,4} &=& c_{10} c_4 c_6 c_7 c_8 + c_{10} c_3 c_6 c_7 c_9 + c_{10} c_2 c_6 c_8 c_9 + 
 c_1 c_{10} c_7 c_8 c_9 + c_5 c_6 c_7 c_8 c_9 \;, \nn \\
a_{4,3} &=&  c_{10} c_3 c_4 c_6 c_7 + c_{10} c_2 c_4 c_6 c_8 + c_1 c_{10} c_4 c_7 c_8 + c_4 c_5 c_6 c_7 c_8 +
  c_{10} c_2 c_3 c_6 c_9 \nn \\
  & &+ c_1 c_{10} c_3 c_7 c_9 + c_3 c_5 c_6 c_7 c_9 + 
 c_1 c_{10} c_2 c_8 c_9 + c_2 c_5 c_6 c_8 c_9 + c_1 c_5 c_7 c_8 c_9\;, \nn \\
a_{3,2} &=&  c_{10} c_2 c_3 c_4 c_6 + c_1 c_{10} c_3 c_4 c_7 + c_3 c_4 c_5 c_6 c_7 + c_1 c_{10} c_2 c_4 c_8 +
  c_2 c_4 c_5 c_6 c_8 \nn \\
  & &+ c_1 c_4 c_5 c_7 c_8 + c_1 c_{10} c_2 c_3 c_9 + c_2 c_3 c_5 c_6 c_9 +
  c_1 c_3 c_5 c_7 c_9 + c_1 c_2 c_5 c_8 c_9\;, \nn \\
a_{2,1} &=& c_1 c_{10} c_2 c_3 c_4 + c_2 c_3 c_4 c_5 c_6 + c_1 c_3 c_4 c_5 c_7 + c_1 c_2 c_4 c_5 c_8 + 
 c_1 c_2 c_3 c_5 c_9 \;, \nn \\
a_{1} &=& c_1 c_2 c_3 c_4 c_5 \;. \label{aIsol11111}
\end{eqnarray}
A solution to the tracelessness constraint is
\begin{eqnarray}
c_7 &=& \delta \alpha_1 \;, \nn \\
c_8 &=& \delta \alpha_2 \;, \nn \\
c_9 &=& \delta \alpha_3 \;, \nn \\
c_{10} &=& \delta \alpha_4 \;, \nn \\
c_6 &=& \delta \alpha_1 \alpha_2 \alpha_3 \alpha_4 \;, \nn \\
c_1 &=& -\left(c_5 \alpha_1\alpha_2\alpha_3 + c_4 \alpha_1\alpha_2\alpha_4 + c_3 \alpha_1\alpha_3\alpha_4 + c_2 \alpha_2\alpha_3\alpha_4\right)\;.
\end{eqnarray}
There are many intersections which must vanish for this solution to hold generally, most notably $\alpha_i \cdot \alpha_j$. Here, because of the strong constraints on intersection numbers, the discussion in section \ref{sec:221split} regarding the fact that the solution is not the most general one possible becomes even more crucial. We proceed with analysing the solution presented since for the purposes of this paper it serves as a useful illustration of the general procedure, but keep in mind that studying more general solutions to $a_{5,4}=0$ in this case is of great importance.

Within the patch $e_1=e_2=e_4=1$ we have that the Tate polynomial can be written as (\ref{tatesing}) with 
\begin{eqnarray}
Y_1 &=& c_2 t + \alpha_1 \delta e_0 z\;, \nn \\
Y_2 &=& c_3 t + \alpha_2 \delta e_0 z\;, \nn \\
Y_3 &=& c_4 t + \alpha_3 \delta e_0 z\;, \nn \\
Y_4 &=& c_5 t + \alpha_4 \delta e_0 z\;, \nn \\
Y_5 &=& -\alpha_2 \alpha_3 \alpha_4 c_2 t - \alpha_1 \alpha_3 \alpha_4 c_3 t - \alpha_1 \alpha_2 \alpha_4 c_4 t - 
   \alpha_1 \alpha_2 \alpha_3 c_5 t + \alpha_1 \alpha_2 \alpha_3 \alpha_4 \delta e_0 z\;, \nn \\
X &=& t^2 - x \;, \nn \\
Q &=& e_3 x^2 - \alpha_2 \alpha_3 \alpha_4 c_2^2 c_3 c_4 c_5 t^3 z - 
 \alpha_1 \alpha_3 \alpha_4 c_2 c_3^2 c_4 c_5 t^3 z - \alpha_1 \alpha_2 \alpha_4 c_2 c_3 c_4^2 c_5 t^3 z - 
 \alpha_1 \alpha_2 \alpha_3 c_2 c_3 c_4 c_5^2 t^3 z 
 \nn \\ & &
 - \alpha_2 \alpha_3 \alpha_4 c_2^2 c_3 c_4 c_5 t x z - 
 \alpha_1 \alpha_3 \alpha_4 c_2 c_3^2 c_4 c_5 t x z - \alpha_1 \alpha_2 \alpha_4 c_2 c_3 c_4^2 c_5 t x z - 
 \alpha_1 \alpha_2 \alpha_3 c_2 c_3 c_4 c_5^2 t x z -  
 \nn \\ & &
 \alpha_2 \alpha_3 \alpha_4^2 c_2^2 c_3 c_4 \delta e_0 t^2 z^2 - 
 \alpha_1 \alpha_3 \alpha_4^2 c_2 c_3^2 c_4 \delta e_0 t^2 z^2 - 
 \alpha_1 \alpha_2 \alpha_4^2 c_2 c_3 c_4^2 \delta e_0 t^2 z^2 - 
 \alpha_2 \alpha_3^2 \alpha_4 c_2^2 c_3 c_5 \delta e_0 t^2 z^2 -  
 \nn \\ & &
 \alpha_1 \alpha_3^2 \alpha_4 c_2 c_3^2 c_5 \delta e_0 t^2 z^2 - 
 \alpha_2^2 \alpha_3 \alpha_4 c_2^2 c_4 c_5 \delta e_0 t^2 z^2 - 
 3 \alpha_1 \alpha_2 \alpha_3 \alpha_4 c_2 c_3 c_4 c_5 \delta e_0 t^2 z^2 - 
 \nn \\ & &
 \alpha_1^2 \alpha_3 \alpha_4 c_3^2 c_4 c_5 \delta e_0 t^2 z^2 - 
 \alpha_1 \alpha_2^2 \alpha_4 c_2 c_4^2 c_5 \delta e_0 t^2 z^2 - 
 \alpha_1^2 \alpha_2 \alpha_4 c_3 c_4^2 c_5 \delta e_0 t^2 z^2 - 
 \alpha_1 \alpha_2 \alpha_3^2 c_2 c_3 c_5^2 \delta e_0 t^2 z^2 - 
 \nn \\ & &
 \alpha_1 \alpha_2^2 \alpha_3 c_2 c_4 c_5^2 \delta e_0 t^2 z^2 - 
 \alpha_1^2 \alpha_2 \alpha_3 c_3 c_4 c_5^2 \delta e_0 t^2 z^2 - 
 \alpha_2 \alpha_3 \alpha_4^2 c_2^2 c_3 c_4 \delta e_0 x z^2 - 
 \alpha_1 \alpha_3 \alpha_4^2 c_2 c_3^2 c_4 \delta e_0 x z^2 - 
 \nn \\ & &
 \alpha_1 \alpha_2 \alpha_4^2 c_2 c_3 c_4^2 \delta e_0 x z^2 - 
 \alpha_2 \alpha_3^2 \alpha_4 c_2^2 c_3 c_5 \delta e_0 x z^2 - 
 \alpha_1 \alpha_3^2 \alpha_4 c_2 c_3^2 c_5 \delta e_0 x z^2 - 
 \alpha_2^2 \alpha_3 \alpha_4 c_2^2 c_4 c_5 \delta e_0 x z^2 - 
 \nn \\ & &
 3 \alpha_1 \alpha_2 \alpha_3 \alpha_4 c_2 c_3 c_4 c_5 \delta e_0 x z^2 - 
 \alpha_1^2 \alpha_3 \alpha_4 c_3^2 c_4 c_5 \delta e_0 x z^2 - 
 \alpha_1 \alpha_2^2 \alpha_4 c_2 c_4^2 c_5 \delta e_0 x z^2 - 
 \alpha_1^2 \alpha_2 \alpha_4 c_3 c_4^2 c_5 \delta e_0 x z^2 - 
 \nn \\ & &
 \alpha_1 \alpha_2 \alpha_3^2 c_2 c_3 c_5^2 \delta e_0 x z^2 - 
 \alpha_1 \alpha_2^2 \alpha_3 c_2 c_4 c_5^2 \delta e_0 x z^2 - 
 \alpha_1^2 \alpha_2 \alpha_3 c_3 c_4 c_5^2 \delta e_0 x z^2 - 
 \alpha_2 \alpha_3^2 \alpha_4^2 c_2^2 c_3 \delta^2 e_0^2 t z^3 - 
 \nn \\ & &
 \alpha_1 \alpha_3^2 \alpha_4^2 c_2 c_3^2 \delta^2 e_0^2 t z^3 - 
 \alpha_2^2 \alpha_3 \alpha_4^2 c_2^2 c_4 \delta^2 e_0^2 t z^3 - 
 2 \alpha_1 \alpha_2 \alpha_3 \alpha_4^2 c_2 c_3 c_4 \delta^2 e_0^2 t z^3 - 
 \alpha_1^2 \alpha_3 \alpha_4^2 c_3^2 c_4 \delta^2 e_0^2 t z^3 - 
 \nn \\ & &
 \alpha_1 \alpha_2^2 \alpha_4^2 c_2 c_4^2 \delta^2 e_0^2 t z^3 - 
 \alpha_1^2 \alpha_2 \alpha_4^2 c_3 c_4^2 \delta^2 e_0^2 t z^3 - 
 \alpha_2^2 \alpha_3^2 \alpha_4 c_2^2 c_5 \delta^2 e_0^2 t z^3 - 
 2 \alpha_1 \alpha_2 \alpha_3^2 \alpha_4 c_2 c_3 c_5 \delta^2 e_0^2 t z^3 - 
 \nn \\ & &
 \alpha_1^2 \alpha_3^2 \alpha_4 c_3^2 c_5 \delta^2 e_0^2 t z^3 - 
 2 \alpha_1 \alpha_2^2 \alpha_3 \alpha_4 c_2 c_4 c_5 \delta^2 e_0^2 t z^3 - 
 2 \alpha_1^2 \alpha_2 \alpha_3 \alpha_4 c_3 c_4 c_5 \delta^2 e_0^2 t z^3 - 
 \alpha_1^2 \alpha_2^2 \alpha_4 c_4^2 c_5 \delta^2 e_0^2 t z^3 - 
 \nn \\ & &
 \alpha_1 \alpha_2^2 \alpha_3^2 c_2 c_5^2 \delta^2 e_0^2 t z^3 - 
 \alpha_1^2 \alpha_2 \alpha_3^2 c_3 c_5^2 \delta^2 e_0^2 t z^3 - 
 \alpha_1^2 \alpha_2^2 \alpha_3 c_4 c_5^2 \delta^2 e_0^2 t z^3 - 
 \alpha_2^2 \alpha_3^2 \alpha_4^2 c_2^2 \delta^3 e_0^3 z^4 - 
 \nn \\ & &
 \alpha_1 \alpha_2 \alpha_3^2 \alpha_4^2 c_2 c_3 \delta^3 e_0^3 z^4 - 
 \alpha_1^2 \alpha_3^2 \alpha_4^2 c_3^2 \delta^3 e_0^3 z^4 - 
 \alpha_1 \alpha_2^2 \alpha_3 \alpha_4^2 c_2 c_4 \delta^3 e_0^3 z^4 - 
 \alpha_1^2 \alpha_2 \alpha_3 \alpha_4^2 c_3 c_4 \delta^3 e_0^3 z^4 - 
 \nn \\ & &
 \alpha_1^2 \alpha_2^2 \alpha_4^2 c_4^2 \delta^3 e_0^3 z^4 - 
 \alpha_1 \alpha_2^2 \alpha_3^2 \alpha_4 c_2 c_5 \delta^3 e_0^3 z^4 - 
 \alpha_1^2 \alpha_2 \alpha_3^2 \alpha_4 c_3 c_5 \delta^3 e_0^3 z^4 - 
 \alpha_1^2 \alpha_2^2 \alpha_3 \alpha_4 c_4 c_5 \delta^3 e_0^3 z^4 - 
 \nn \\ & &\alpha_1^2 \alpha_2^2 \alpha_3^2 c_5^2 \delta^3 e_0^3 z^4 \;.
\end{eqnarray}

\section{Relation to other approaches to \texorpdfstring{$U(1)$}{U(1)}s}
\label{sec:relother}
\subsection{Relation to the \texorpdfstring{$U(1)$}{U(1)}-restricted Tate model}
\label{sec:reltate}

In \cite{Grimm:2010ez} a method for constructing elliptic fibrations that support a global $U(1)$ symmetry was proposed. In this appendix we discuss the relation of this method to the results discussed in this paper. The model of \cite{Grimm:2010ez} corresponds to the $4 - 1$ factorisation but with the added constraint that $c_1=1$ \cite{Marsano:2009gv} so that the {\bf 10}-matter curve $C_{{\bf 10}^{(2)}}$ in (\ref{U1X10curve}) is switched off. It was shown that after an appropriate coordinate transformation 
\begin{equation}
x \rightarrow \tilde{x} + \left(w c_0 z\right)^2 \;,\;\; y \rightarrow \tilde{y} - \left(w c_0 z\right)^3 \;,
\end{equation} 
the monomial associated to $a_6$ in (\ref{tateform}) vanishes. In that case it was argued using Tate's algorithm that after the transformation there is an $SU(2)$ singularity over the curve
\begin{equation}
\tilde{a}_{4,3} = \tilde{a}_{3,2} = 0 \;. \label{tatesu2}
\end{equation} 
This singularity can then be resolved by a blow-up $x  \rightarrow x s, y \rightarrow y s$ which accounts for the additional $U(1)$ (see also \cite{Krause:2011xj,Grimm:2011fx}). The $SU(2)$ singularity was also identified in a different way in \cite{Braun:2011zm} by moving to the Sen coordinates in which case the Weierstra\ss{} polynomial takes the form
\begin{equation}
Y_- Y_+ - X Q = z^6 a_6 \;, \label{senconi}
\end{equation} 
and so in a coordinate basis where $a_6=0$ takes the form of a conifold. The singularity locus in the Sen coordinates $Y_+=Y_-=X=Q=0$ coincided with locus determined from Tate's algorithm $\tilde{x}=\tilde{y}=\tilde{a}_{4,3} = \tilde{a}_{3,2} = 0$.

It is possible to generalise this approach to understanding the $U(1)$s to other cases as follows. We consider the special case where the $U(1)$ is associated to a section which can be written in the form $x=A^2$ and $y=-A^3$, with $A$ and $B$ being some holomorphic polynomials. Since we have been discussing sections that satisfy $y^2=x^3$ the constraint is that the additional holomorphic equation specifying the section can be written in the form $x=A^2$. If this is possible, then the procedure employed in the $U(1)$-restricted model can be applied generally. The idea is to shift the coordinates by the section $x \rightarrow \tilde{x}+A^2$, $y \rightarrow \tilde{y}-A^3$ and in the new coordinates it must be that $\tilde{a}_6=0$ since at $\tilde{x}=\tilde{y}=0$ we recover the section that satisfies $P_T= \tilde{a}_6 z^6 = 0$. Once this coordinate choice is made the singularity can be identified using the two methods described above.

Therefore the particular case studied in \cite{Grimm:2010ez} was applying this procedure to a fibration with a section satisfying the above constraints with $A=c_0 w z$. This section is related to the general section for the $4 - 1$ case which is identified in (\ref{ptf41}) as one of the factors $Y_i$ to be
\begin{equation}
A c_1 = c_0 w z \;,
\end{equation} 
where we parametrically solved the $y^2=x^3$ part of the section by setting $t = y/x = - A$. Now we see that $A$ is only holomorphic if we set $c_1=1$ and so turning off one of the ${\bf 10}$-matter curves was crucial to the success of the procedure. Generally however $A$ is only meromorphic and diverges on the second ${\bf 10}$-matter curve and where this procedure breaks down. For this more general case the approach described in this paper must be adopted. It is possible to check that if we continue with the $U(1)$-restricted procedure without worrying about the meromorphicity in $c_1$ the singularity locus identified using Tate's algorithm (\ref{tatesu2}) or using the Sen coordinates (\ref{senconi}) both match the singularity locus obtained using our procedure (\ref{ptf41}).

\subsection{Relation to split spectral cover models}
\label{App-SCC}

In the local limit the split Tate model flows to the split spectral cover construction. The local limit is well defined before the resolution of the $SU(5)$ singularity and corresponds to taking $w \rightarrow 0$. The local limit of the Tate model (\ref{tpartsec}) was studied in \cite{Marsano:2010ix,Marsano:2011hv} where the section (\ref{tatesection}) was termed the Tate divisor. In order to recover the spectral cover the limit must be taken such that also $t \rightarrow 0$ while keeping the ratio finite \cite{Marsano:2010ix}
\begin{equation}
w \rightarrow 0 \;,\;\; \frac{w}{t} \rightarrow s \;, \label{loclim}
\end{equation} 
in the patch $z=1$. After the proper transform of dividing out by the overall factor of $t^5$ we recover the Higgs bundle on $S_{GUT}$
\begin{equation}
b_5 + b_4 s + b_3 s^2 + b_2 s^3 + b_0 s^5 = 0 \;. \label{higgsbuna4}
\end{equation} 
Here we denote
\begin{eqnarray}
\left.a_1\right|_{w=0} = b_5 \;, \;
\left.a_{2,1}\right|_{w=0} = b_4 \;, \;
\left.a_{3,2}\right|_{w=0} = b_3 \;, \;
\left.a_{4,3}\right|_{w=0} = b_2 \;, \;
\left.a_{6,5}\right|_{w=0} = b_0 \;.
\end{eqnarray}
We should think of this in terms of an underlying $E_8$ symmetry broken according to $E_8 \rightarrow SU(5)_{GUT} \times SU(5)_{\perp}$ in two equivalent ways. Either through an 8-dimensional gauge theory on the GUT brane with gauge group $E_8$ that is broken to $SU(5)_{GUT}$ by a spatially varying adjoint Higgs field $\varphi$ with vev in the $SU(5)_{\perp}$, the precise map being \cite{Donagi:2009ra} 
\begin{equation}
b_1 = {\mathrm Tr}\left[\varphi \right] \;,\; b_2 = -\frac12 {\mathrm Tr}\left[\varphi^2 \right] \;,\; b_5 = \det\left[\varphi \right] \;. \label{higgsbun}
\end{equation} 
The other way is through an $A_4$ singularity, corresponding to $SU(5)_{\perp}$, that is fibred over $S_{GUT}$ \cite{Beasley:2008dc}. A fully deformed $A_4$ singularity takes the form
\begin{equation}
y^2 = x^2 + \prod_{i=1}^5 \left(s+t_i\right) \;,
\end{equation} 
where the $t_i$ are 5 deformation parameters, which are functions on $S_{GUT}$, that can be explicitly mapped to the Cartan $U(1)$s inside $SU(5)_{\perp}$. The proper identification with the $b_i$ is simply the expansion
\begin{equation}
\prod_{i=1}^5 \left(s+t_i\right) = \left(\frac{b_5}{b_0}\right) + \left(\frac{b_4}{b_0}\right) s + \left(\frac{b_3}{b_0}\right) s^2 + \left(\frac{b_2}{b_0}\right) s^3 + s^5 \;. \label{splita4}
\end{equation} 
This determines the $b_i$ as the elementary symmetric polynomials in the $t_i$. 

The $A_4$ singularity has a Weyl group action which interchanges the $t_i$ so as to preserve the $b_i$. Generally the fibration over $S_{GUT}$ can act with this group which in F-theory is termed monodromies \cite{Hayashi:2009ge,Donagi:2009ra}. More generally we can think of the Higgs bundle as taking value in various subgroups of $SU(5)_{\perp}$ that preserve some $U(1)$ symmetries, and in diagonalising the Higgs so that the map (\ref{higgsbun}) holds, branch cuts are induced in the form of the $t_i$ as functions on $S_{GUT}$ which map them to each other as we move around the branch \cite{Cecotti:2010bp}. The case where the Higgs preserves the full Cartan of $SU(5)_{\perp} \supset S\left[U(1)^5\right]$ is mapped to the case where there are no monodromies, while maintaining smaller Abelian subgroups corresponds to non-trivial monodromies. 

This maps directly to the product structure of (\ref{splita4}), where we see that under no identification of the $t_i$, (\ref{higgsbuna4}) factorises into 5 factors. Each factor corresponds to a $U(1)$ with a tracelessness constraint $b_1=0$ leaving the 4 Cartan $U(1)$s as linearly independent. As we identify the $t_i$ (\ref{higgsbuna4}) decomposes into fewer factors implying fewer $U(1)$s and finally if the fibration uses the full Weyl group there is no splitting at all and no $U(1)$s.

Exactly this structure is what is termed a split spectral cover, where we simply compactify the surface (\ref{splita4}) by writing $s$ in terms of homogeneous coordinates $s= U/V$ \cite{Donagi:2009ra}.

The discussion presented is the local understanding of the required splitting structure of (\ref{higgsbuna4}) in order to preserve a $U(1)$. So for example the $4-1$ factorisation is such that 
\begin{equation}
\left(c_0 s + c_1 \right) \left(s^4 d_0 + s^3 d_1 + s^2 d_2 + s d_3 + d_4 \right) = b_5 + b_4 s + b_3 s^2 + b_2 s^3 + b_0 s^5 \;, \label{localsplit41}
\end{equation} 
where the $c_i$ and $d_i$ are holomorphic functions on $S_{GUT}$. This fixes the form of the $b_i$ and imposes a tracelessness constraint on the $c_i$ and $d_i$.

Note that, of course, the local splitting is a weaker constraint on the full $b_i$ than a factorised Tate model which also constrains the $w$ dependence of the $b_i$. For example, it was shown in \cite{Choi:2012pr} that for the case of a Heterotic dual there are constraints on the complex structure moduli of the F-theory CY which manifest in the specific form of the $w$ dependence of the $b_i$.  These higher order terms in the $b_i$ precisely take the form so as to respect the appropriate factorisation structure (\ref{ptf41}) which means they can be written in terms of higher order terms in the $c_i$ and $d_i$, specifically
\begin{equation}
d_0 = \left.d_0\right|_{w=0} - F w c_0 \;,\;\; d_1 = \left.d_1\right|_{w=0} + F w c_1  \;,
\end{equation} 
where $F$ is some arbitrary function.

\end{document}